\documentclass[a4paper,11pt]{article}
\usepackage[utf8]{inputenc}
\usepackage[english]{babel}
\usepackage{braket}
\usepackage{xspace}
\usepackage{bbm}
\usepackage{mathdots}
\usepackage{stackrel}
\usepackage{xcolor}
\usepackage{mathtools}
\usepackage{bbm}
\usepackage{subfigure}
\usepackage[T1]{fontenc}               
\usepackage[left=3cm,
            right=2.5cm,
            top=2.5cm,
            bottom=3cm
            ]{geometry}                
\usepackage{graphicx}
\usepackage{mathrsfs}
\usepackage{appendix}
\usepackage{fancyhdr}
\usepackage{amsmath,                   
            amssymb,                   
            amsthm}                    
\usepackage{cite}

\usepackage{setspace} 

\usepackage[colorlinks=true,linkcolor=blue, citecolor=blue, bookmarks]{hyperref}

\numberwithin{equation}{section}

\def \be {\begin{equation}} 
\def \ee {\end{equation}} 
\def \l {\left(} 
\def \r {\right)} 
\def \la {\langle} 
\def \ra {\rangle}  

\date{}

\title{R\'enyi entropy and negativity for massless Dirac fermions at conformal interfaces and junctions}

\author{Luca Capizzi$^1$, Sara Murciano$^1$, and Pasquale Calabrese$^{1,2}$}
\begin{document}
\maketitle

$^1$SISSA and INFN Sezione di Trieste, via Bonomea 265, 34136 Trieste, Italy.\\
$^{2}$International Centre for Theoretical Physics (ICTP), Strada Costiera 11, 34151 Trieste, Italy.\\
\begin{abstract} 
We investigate the ground state of a (1+1)-dimensional conformal field theory built with $M$ species of massless free Dirac fermions coupled at one boundary point via a 
conformal junction/interface. Each CFT represents a wire of finite length $L$.
We develop a systematic strategy to compute the R\'enyi entropies for a generic bipartition between the wires and the entanglement negativity 
between two non-complementary sets of wires. 
Both these entanglement measures turn out to grow logarithmically with $L$ with an exactly calculated universal prefactor depending on the details of the junction 
and of the bipartition. 
These analytic predictions are tested numerically for junctions of free Fermi gases, finding perfect agreement.
\end{abstract}

 \tableofcontents

\section{Introduction}
The entanglement content of extended quantum systems has been investigated in the last two decades in many different contexts, ranging from condensed matter 
\cite{intro2,eisert-2010,afov-08,Laflorencie-16} to high energy and black hole physics \cite{bkls-86,Srednicki-93,Raamsdonk,maldacena,rt-06}. 
The most successful way to quantify the many-body entanglement is via the R\'enyi entropies: 
given a system in a pure state $|\Phi\rangle$ and a bipartition $A \cup B$, the subsystem $A$ is described by the reduced density matrix 
$\rho_A=\mathrm{Tr}_B |\Phi\rangle\langle \Phi|$, 
and the the R\'enyi entropies are  
\begin{equation}\label{eq:renyidef}
    S_n(A)=\frac{1}{1-n}\log \mathrm{Tr}\l\rho_A^n\r.
\end{equation}
For $n\to 1$ this definition gives the von Neumann entropy $S(A) = - \text{Tr}\l \rho_A \log \rho_A \r$, often called just {\it entanglement entropy}. 
One of the most important use of the R\'enyi entropies has been the characterisation of critical one-dimensional systems:
the distinctive feature is the logarithmic divergence of the entanglement entropy with the (sub)system size and conformal field theory (CFT) provides universal predictions 
for the prefactor of such logarithm.
For example, the vacuum R\'enyi entanglement entropy of an interval $A$ of length $\ell$ embedded in an infinite system is given by \cite{hlw-94,cc-04,cc-09}
\be
S_n(A) =  \frac{c}{6}\Big(1+\frac1n \Big) \log \frac{\ell}{\varepsilon}+\dots,
\label{eq:VNentropyInf}
\ee
where $c$ is the central charge and $\varepsilon$ is a ultraviolet (UV) cutoff. 
This remarkable scaling behaviour is altered at leading order by the presence of a boundaries \cite{cc-04}.
For conformally invariant boundary conditions (bc's), the entanglement entropy can be studied via boundary CFT \cite{bcft,cardy-84,c-04}, a framework 
that already found a large number of applications in condensed matter and particle physics, such as quantum impurity problems \cite{s-98}, 
the multi-channel Kondo problem \cite{kondo}, D-brane physics \cite{p-96} etc.
For a finite size CFT of length $2L$ with conformal invariant bc's at the two edges, 
the R\'enyi entanglement entropy between the half-chain $A = [0,L]$ and the other half is \cite{cc-04,cc-09}
\be 
S_n(A) = \frac{c}{12}\Big(1+\frac1n \Big)\log \frac{L}{\varepsilon} + \dots,
\label{eq:VNentropyFin}
\ee
up to finite terms that depend on the bc's. 
At leading order in $L/\varepsilon\rightarrow \infty$, Eqs. \eqref{eq:VNentropyInf} and \eqref{eq:VNentropyFin} differ by a factor $2$. 
This is heuristically understood because the two geometries differ by the number of entangling points and, in general, one expects the entanglement entropy to be 
proportional to the size of the boundary of the subsystem. 
In both geometries mentioned above the origin of the entanglement relies on the presence of {\it completely transmissive entangling points}, 
resulting in some degree of quantum coherence among the subsystem and its complement. 
Conversely, when the entangling points are completely reflective because of  some additional boundary conditions, the subsystems decouple, 
and the entanglement entropy between them vanishes.

A natural generalisation of the above scenarios regards the intermediate setting in which the entangling points are partially transmitting and reflecting \cite{bddo-02}. 
In the literature, such special situations are known as permeable interfaces, defects, or impurities (and indeed we will refer to them using all these equivalent names). 
A crucial result is that for free massless theories the defect is marginal \cite{kf-92} and so can alter the leading behaviour of the entanglement entropy.
Conversely, interactions make the defect either relevant or irrelevant \cite{kf-92} ending up asymptotically in a completely reflective or transmitting situation, 
respectively, as shown also by the scaling of the entanglement entropy itself \cite{coc-13,p-051}.
For free theories in the presence of a conformal interface (i.e., scale invariant)  Sakai and Satoh exploited boundary CFT to show that the 
scaling of the entanglement entropy, \eqref{eq:VNentropyFin} for $n=1$, is modified as \cite{ss-08}
\be
S(A) = \frac{c_{\text{eff}}(\sqrt{\mathcal{T}})}{6}\log \frac{L}{\varepsilon} + \dots,
\label{eq:VNentropySS}
\ee
where $\mathcal{T}$ is a parameter which represents the transmission probability and $c_{\text{eff}}(\sqrt{\mathcal{T}})$, dubbed as \textit{effective central charge}, is 
a monotonic function of its argument satisfying
\be
c_{\text{eff}}(0)=0, \quad c_{\text{eff}}(1) = c.
\ee
Ref. \cite{ss-08} focuses on the free massless boson, but Eq. \eqref{eq:VNentropySS} with a different $c_{\text{eff}}(\sqrt{\mathcal{T}})$ has been subsequently derived also for 
free massless fermions both by means of CFT \cite{bb-15,bbr-13,tm-21}, and explicitly solving microscopic models \cite{cmv-12,ep-12,p-05,ep-10,ep2-12} in the same 
universality class.
While the scaling in Eq. \eqref{eq:VNentropySS} is expected to be a generic feature of conformal invariant (1+1)-dimensional systems, the explicit functional form of the effective central charge depends both on the theory and the details of the interface, eventually encoded in a interface operator (or, equivalently, in a boundary 
state as explained in \cite{bb-15}).
We mention that a class of completely transmissive interfaces, dubbed \text{topological interfaces}, has been also considered in the literature \cite{ffrs-04,ffrs-07,aamf-16,bbjs-16,Jaud-2017,rs-22,rpr-22}. 
While their effective central charge is always $c$, and they could be erroneously considered trivial, the $O(1)$ terms shrugged off 
in Eq. \eqref{eq:VNentropySS} still contains important information about the boundary conditions, strictly related to the boundary entropy of Affleck-Ludwig \cite{al-91}.

\begin{figure}[t]
\centering
	\includegraphics[width=0.6\linewidth]{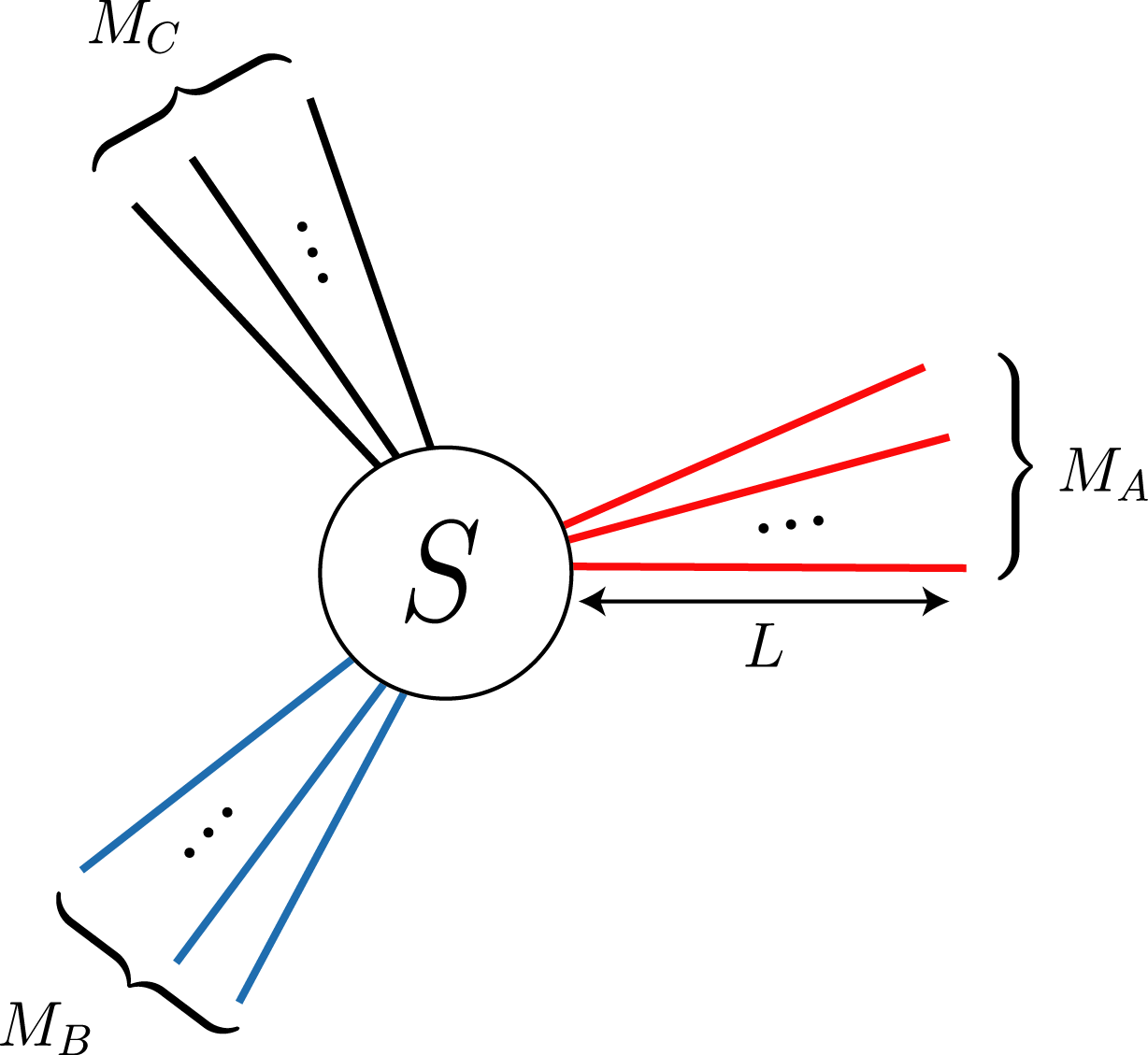}
    \caption{The conformal junction: $M$ wires are joined together at $x=0$ by a conformally invariant scattering matrix $S$. 
    We consider a tripartition in three sets $A, B, C$ with $M_A,M_B, M_C$ wires each. 
    The entanglement between $A$ and $B$ is given by the negativity \eqref{negdef}. 
    A bipartite configuration is simply obtained by letting $M_C=0$.}
    \label{fig:star}
\end{figure}

The permeable interface between two CFTs can be generalised to a junction of $M$ wires.
The resulting geometry is depicted in Fig. \ref{fig:star} in which the junction is fully characterised by a scattering matrix $S$ between the wires. 
Imposing that this matrix $S$ preserves conformal invariance, one finds consistency conditions that have been studied and solved for a large number of physical 
configurations \cite{nfll-99,coa-03,coa-06,bm-06,bms-07,hc-08,bcm-09,bms-09,cmr-13}.
The bipartite entanglement in these conformal (or star) junctions has been studied  in Refs. \cite{cmv-11,cmv-12,cmv-12a,gm-17} but focusing on 
the entanglement between a single wire and the remaining $M-1$ ones. 
A unifying framework to compute the entanglement of a generic bipartition among the wires of the junctions is still missing. 

The conformal junction is also a very obvious setup for the study of multipartite entanglement because it is made of several wires and it is very natural to 
wonder about the entanglement between a subset of them, not only two complementary subsystems.
In this respect, the first configuration that comes to mind is the tripartition in  $A, B, C$ with $M_A,M_B, M_C$ wires each, as depicted in Fig. \ref{fig:star}.
To study the entanglement of this tripartition, one can integrate out the $M_C$ wires in C, to get the reduced density matrix $\rho_{A\cup B}$.
Then the entanglement between $A$ and $B$ with the mixed density matrix $\rho_{A\cup B}$ is measured by the negativity \cite{vidal,plenio-2005}
\be 
\mathcal{E}=\log ||\rho_{A\cup B}^{T_B}||,
\label{negdef}
\ee
where $T_B$ denotes the partial transposition with respect to the degrees of freedom in $B$ and $|| \cdot ||$ stands for the trace norm.
The negativity in the presence of a defect has been computed for a bipartite geometry with $M=2$\cite{ge-20}, exploiting its relation with
the $1/2$-R\'enyi entropy for the bipartition of a pure state, but for a genuinely tripartite geometry at a junction there are no results yet. 

The main goal of this manuscript is to provide a general framework to deal with the entanglement through permeable junctions of $M$ (1 + 1)-dimensional free-fermion CFT. 
Following Refs. \cite{ss-08,bddo-02}, the strategy is to constrain the form of the general boundary state in a folded theory. 
Then, being the theory free, we can reduce the problem to the computation 
of a charged partition function in the presence of this boundary state. 
This approach also allows us to compute the negativity in a tripartite geometry by properly implementing a partial transpose operation for free fermions.

The paper is organised as follows. In section \ref{sec:CFTapproach} we review the folding trick which turns the problem of constructing conformal interfaces into the one of building boundary states. We review the construction of  fermionic boundary states and we compute the partition functions in the junction geometry. 
Using this result and the replica trick, we obtain the entanglement entropy analytically for a generic bipartition between wires. 
In Section \ref{sec:negativity}, we combine the previous formalism with the replica trick for the fermionic negativity. 
This allows us to obtain an analytic prediction that we benchmark against numerical computations in Section \ref{sec:schroedinger}. 
In the same section, we also describe an alternative technique for the computation of the entanglement of a fermion gas on a star graph modelling the junction of interest. 
We draw our conclusions in section \ref{sec:conlusion} and we relegate some technical material about our computations in the Appendix \ref{app1}.

\section{CFT approach: description of the method and the application to the R\'enyi entropy}\label{sec:CFTapproach}

 \begin{figure}[t]
\centering
	\includegraphics[width=0.82\linewidth]{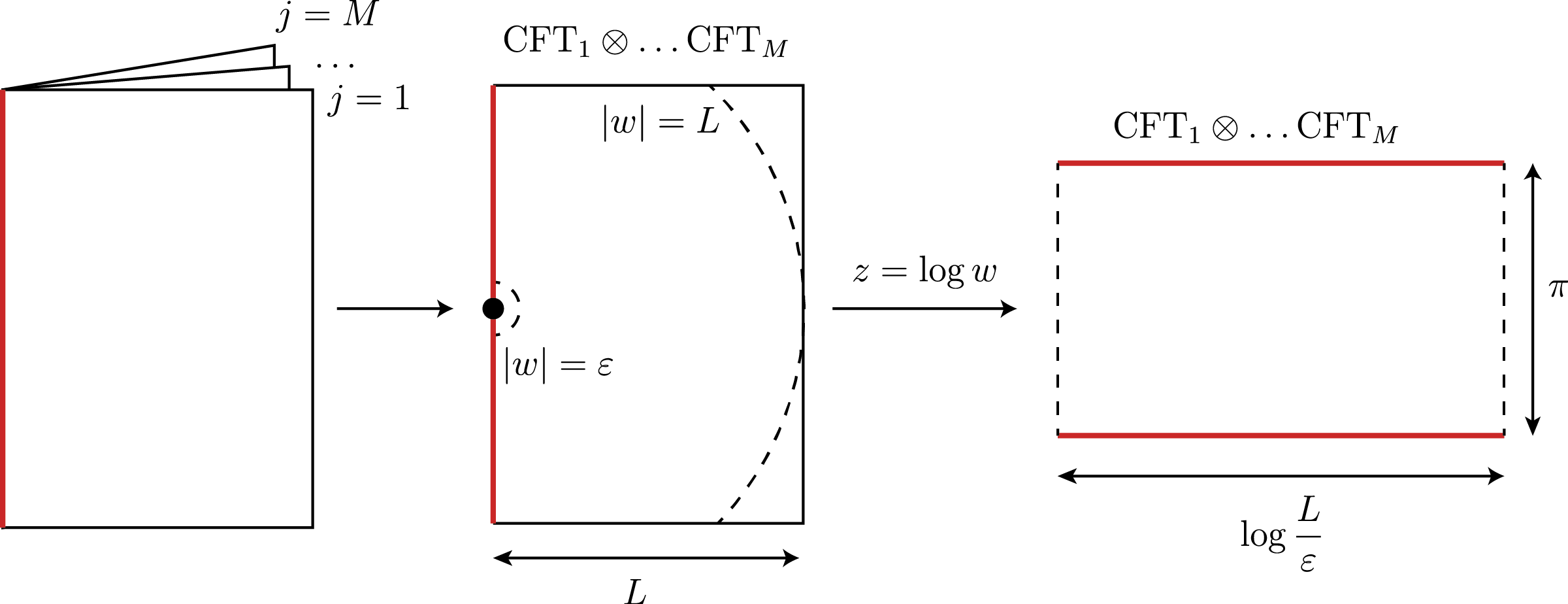}
    \caption{The folding procedure. The junction in Euclidean spacetime is the booklet with the CFTs bound in $x=0$ (left panel).
    The folding consists in merging together the $M$ CFTs in a worldsheet being a single infinite strip with an appropriate boundary state $|S\ra$ at $x=0$ (middle panel).
    To compute the entanglement, we cut the system for $|w|<\varepsilon$ and $w>|L|$ (dashed lines) and map the M-CFT onto a rectangle of size $\log \frac{L}\varepsilon\times \pi$  
    (right) with ${\rm Re}(z) \in [\log \varepsilon,\log L]$ and ${\rm Im}(z) \in [-\pi/2,\pi/2]$. }
    \label{fig:Fold_picture}
\end{figure}

In this section we present the CFT approach for the evaluation of the entanglement in permeable junctions of (1+1)-dimensional free-fermion CFTs, following closely Refs. \cite{bddo-02,ss-08,bb-15}. As a first application, we employ this method to compute the R\'enyi entropies between an arbitrary number of wires at the junction.

Let us consider $M$ wires of length $L$, each of them described by a CFT denoted by
\be
\text{CFT}_j, \quad j=1,\dots,M.
\ee
In Euclidean space-time, the junction looks like a booklet (with each page corresponding to one CFT) bound along the imaginary axis at $x=0$, 
see the left panel of Fig. \ref{fig:Fold_picture}. 
As custom in this kind of systems, we are going to work in the folded picture in which the system is represented as a single CFT 
\be
\text{M-CFT} = \text{CFT}_1 \otimes \dots \otimes \text{CFT}_M,
\ee
i.e. the world-sheet is a single infinite strip of width $L$ 
(we are dealing with a finite size quantum system at zero temperature, so the space-time coordinate $w$ satisfies ${\rm Re}(w) \in [0,L]$)
where $M$ copies of the CFT live.
See the middle panel of Fig. \ref{fig:Fold_picture} for a pictorial representation.
The joining between the distinct wires is specified by the boundary conditions along the lines
\be
{\rm Re}(w) = 0, \quad {\rm Re}(w) = L.
\ee
We require that the boundary condition at ${\rm Re}(w) = L$ decouples the replicas, and can be thus described by a boundary state $\ket{B}$ factorised as
\be
\ket{B} = \ket{B_1}\otimes \dots \ket{B_M},
\ee
with $\ket{B_j}$ being a boundary state of $\text{CFT}_j$. Instead, we assume that the boundary conditions at ${\rm Re}(w) = 0$ (describing the defect/junction), 
in general couple explicitly distinct wires.  
We denote by $ \ket{S}$ the associated boundary state in M-CFT. 
In the remainder of the manuscript, the precise details of the boundary state $\ket{B}$ appearing at ${\rm Re}(w) = L$ would not matter and so we do not specify more about it. 
The physical motivation is that, as long as it decouples the wires, we do not expect that its features affect (at least at leading order) the correlation properties among distinct wires. In contrast, this is not the case for the boundary state $\ket{S}$, and for this reason we have to be very careful about its characterisation.

In order to have under control ultraviolet  and  infrared  divergences in the entanglement entropy, 
a standard trick \cite{hlw-94,cw-94,ct-16,act-17,ot-15} consists in cutting the theory for $|w| < \varepsilon$ and $|w| > L$ (see Fig. \ref{fig:Fold_picture}, middle panel). 
The cut strip can then be mapped into a rectangle by the conformal transformation
\be
z = \log w.
\ee
The semicircles $|w| = \varepsilon$ and $|w| = L$  are mapped respectively onto the segments
\be
z \in \log \varepsilon + i[-\pi/2,\pi/2], \quad z \in \log L +  i[-\pi/2,\pi/2].
\ee
The defect line at $\text{Re}(w) = 0$ is split into the two lines $\text{Im}(z) = \pm \pi/2$. This mapping is shown  in the right panel of Fig. \ref{fig:Fold_picture}.
%
The partition function in this geometry can be written as
\be
\mathcal{Z} = \bra{S}\exp\l -\pi H \r\ket{S},
\label{ZvsS}
\ee
where $\pi$ is the height of the rectangle (see Fig. \ref{fig:Fold_picture}),
while the hamiltonian is (up to the Casimir energy which does not play any role in our discussion)
\be
H = \frac{2\pi}{\log \frac{L}{\varepsilon}}\l L_0 + \bar{L}_0 \r,
\ee
with $L_0,\bar{L}_0$ being generators of the Virasoro algebra of $\text{CFT}_1\otimes \dots \text{CFT}_M$.

So far, everything is general and no assumption on the bulk theory or the boundary state $\ket{S}$ has been made yet. 
However, the knowledge of $\ket{S}$ is required to evaluate the partition function (and, by replicas, the entanglement). 
From now on, we thus restrict the analysis to massless free fermions for which we can provide a precise characterisation for the boundary state $\ket{S}$.

\subsection{Boundary states for free-fermions}

In this section, we first review the construction of boundary states for a theory of many species of massless Majorana fermions \cite{bddo-02,gm-17}. 
Then we discuss the straightforward generalisation to Dirac fermions, obtained through a doubling of the degrees of freedom \cite{DiFrancesco-97}.

We consider $M$ species of Majorana fermions. This CFT has central charge $c= M/2$ and it is described in terms of the left/right chiral fermionic fields
\be
\psi^j, \quad \bar{\psi}^j, \quad j=1,\dots,M.
\ee
In radial quantisation \cite{DiFrancesco-97}, restricting the analysis to the Neveu-Schwarz (NS) sector, one can decompose the fermionic fields in their Laurent modes
\be
\psi^j(z) = \sum_{k \in \mathbb{Z}+1/2} \frac{\psi^j_{k}}{z^{k+1/2}}, \quad \bar{\psi}^j(\bar{z}) = \sum_{k \in \mathbb{Z}+1/2} \frac{\bar{\psi}^j_{k}}{\bar{z}^{k+1/2}}.
\ee
(In the Ramond sector, $k$ would be integer and the discussion would be slightly more involved due to the presence of a zero mode for $k=0$.)
Within this convention, the creation/annihilation operators of a fermion of the $j$-th species in the mode $k$ ($k>0$) are $\psi^{j}_{\mp k}$. The number $k$ is (proportional to) the momentum of the particle. More precisely, one can show that the commutation relations between the fermionic fields and the Virasoro operators $L_0,\bar{L}_0$ are
\be
[L_0,\psi^j_{-k}] = k\psi^j_{-k}, \quad [\bar{L}_0,\bar{\psi}^j_{-k}] = k\bar{\psi}^j_{-k}.
\ee
The effect of the scattering matrix $S$ at the junction (as in  in Fig. \ref{fig:star}) is nothing but a consistency condition for the state boundary state $\ket{S}$ reading 
\be
\l\psi^j_k + iS_{jj'}\bar{\psi}^{j'}_{-k} \r \ket{S}=0,
\label{cons}
\ee
where, hereafter, repeated indices are summed over. 
It has been shown \cite{bbr-13,bbr-12}, that in order to preserve conformal invariance at the boundary, $S$ must be orthogonal
\be
S \in O(M).
\ee
In particular the possible $k$-dependence of the scattering matrix is ruled out by scale invariance. 
The solution for $\ket{S}$ of Eq. \eqref{cons} is simply 
\be
\ket{S} = \prod_{k \in \mathbb{N}-1/2}\exp\l i S_{jj'}\psi^j_{-k}\bar{\psi}^{j'}_{-k}\r\ket{0},
\ee
with $\ket{0}$ being the vacuum of the theory. Notice that the different values of $k$ are decoupled, a fact that will simplify the forthcoming computations. 
Nevertheless, in general, different species of particles are coupled, due to the possible occurrence of  non-diagonal terms in the matrix $S$. Those terms represent physically the amplitudes of transmission between different wires and cause the entanglement among them.

We now consider a theory of $M$ free Dirac fermions (having central charge $c= M$), for which the associated fields are
\be
\Psi^j, \quad {\Psi^\dagger}^j, \quad \overline{\Psi}^j, \quad {{\overline{\Psi}}^\dagger}^j,\quad  j=1,\dots,M,
\ee
where $\Psi$ and $\Psi^\dagger$ represent the particles/antiparticles respectively. 
This theory is equivalent to a theory with $2M$ Majorana fermions, and so the previous derivation is valid also in this case. 
The number of degrees of freedom is doubled and one should take an orthogonal real $2M\times 2M$ scattering matrix
$S \in O(2M)$.
However, if we further impose that the global $U(1)$ symmetry
\be
\Psi \rightarrow e^{i\theta}\Psi, \quad  \Psi^\dagger \rightarrow e^{-i\theta}\Psi^\dagger
\ee
is preserved by the boundary conditions, there are additional constraints on the scattering matrix. 
This requirement corresponds to the property that a left/right particle can be produced from the vacuum (through the boundary state) together with its right/left antiparticle only.
Requiring that this symmetry is preserved by the boundary conditions, we end up into a complex unitary scattering matrix
\be
S \in U(M),
\ee
that constrains the boundary state $\ket{S}$ as
\be
\l{\Psi^\dagger}^j_k + iS_{jj'}{\overline{\Psi}}^{j'}_{-k} \r \ket{S}=0, \quad \l\Psi^j_k + i\overline{S}_{jj'}{{\overline{\Psi}}^\dagger}^{j'}_{-k} \r \ket{S}=0,
\ee
with $\bar{S}$ being the matrix complex conjugated to $S$. The solution of such constraint is 
\be
\ket{S} = \prod_{k \in \mathbb{N}-1/2}\exp\l i S_{jj'}{\Psi^\dagger}^j_{-k}\overline{\Psi}^{j'}_{-k} + (\Psi \leftrightarrow \Psi^\dagger)\r\ket{0}.
\label{eq_DiracBstate}
\ee
A property of the state $\ket{S}$ in Eq. \eqref{eq_DiracBstate} is that it contains two decoupled contributions, depending on the right and left moving particles. 
We will use this property to simplify the computations in the following sections.

\subsection{R\'enyi entropies for a generic bipartition between wires}

We describe how to compute the $n$-th R\'enyi entropy of a subset made up of $M_A\leq M$ wires via the replica trick. 
Given a subsystem $A$ of a generic QFT, the R\'enyi entropies \eqref{eq:renyidef} of integer order $n$ can be obtained 
in a replicated theory with $n$ copies of the QFT, i.e. in QFT$^{\otimes n}$, which are cyclically joined along $A$
by a branch-cut connecting the $i$-th and the $(i+1)$-th replica \cite{cc-04}. 
The moments of the reduced density matrices can be then written in terms of a ratio of partition functions  as \cite{cc-04,cc-09}
\be
\text{Tr}\l \rho_A^n \r = \frac{\mathcal{Z}_n}{\mathcal{Z}^n_1},
\label{ratioZ}
\ee
where $\mathcal{Z}_n$ is the partition function of the replicated theory while $\mathcal{Z}^n_1$ is just the partition function of a single replica raised to the $n$-th power.

In the case of the bulk free Dirac fermion, the partition function $\mathcal{Z}_n$ can be further factorised using the replica diagonalisation as, e.g., shown in \cite{cfh-05}. 
Within this method, the replicated partition function $\mathcal{Z}_n$ becomes the product of $n$ single-replica $U(1)$ {\it charged} partition functions with 
flux $\alpha_p=\frac{2\pi p}{n}$ ($p=-\frac{n-1}{2},\dots, \frac{n-1}{2}$),  
each of them denoted by $\mathcal{Z}_1(\alpha_p)$. The flux $e^{i\alpha_p}$ is inserted along the branch-cut of $\mathcal{Z}_n$ that
can be rewritten as 
\be
\mathcal{Z}_n = \prod^{\frac{n-1}{2}}_{p =-\frac{n-1}{2}} \mathcal{Z}_1\l \alpha_p = \frac{2\pi p}{n} \r.
\label{eq:prod_flux}
\ee
In Ref. \cite{cfh-05} this factorisation is derived by writing $\mathcal{Z}_n$ as a $2n$-dimensional Gaussian integral, whose diagonalisation leads to the product of $n$ 
two-dimensional Gaussian integrals. 
Plugging Eq. \eqref{eq:prod_flux} for ${\cal Z}_n$ into Eq. \eqref{ratioZ} one has $\text{Tr}\l \rho_A^n \r = \prod_p \l \mathcal{Z}_1 (\alpha_p)/\mathcal{Z}_1\r $ in which the ratio
$\mathcal{Z}_1(\alpha_p)/\mathcal{Z}_1$  can be  expressed as the vacuum expectation value of the operator associated to the action of the $U(1)$ 
symmetry restricted to $A$, namely
\be
\mathcal{Z}_1(\alpha) = \la 0| e^{i\alpha Q_{A}}|0\ra.
\ee
Here $Q_A$ is the charge operator which counts the difference between particles and antiparticles in the subsystem $A$, while $\ket{0}$ is the vacuum of the theory.
Notice that these charged partition sums are the same appearing in the calculation of the symmetry resolved entanglement \cite{gs-18,xas-18,brc-19,mdc-20}.

While Ref. \cite{cfh-05} and most of the subsequent literature focus on the ground state of the system in the absence of boundaries, 
the same considerations apply more generically and in particular to the case of interest here. 
The reason is that the boundary state of interest \eqref{eq_DiracBstate} is Gaussian (it is  an exponential of a bilinear of fermions) 
and thus the functional measure is Gaussian too: in other words, the theory is free both in the bulk and at the boundary (we stress that this property cannot be assumed a priori 
for a quadratic bulk theory, because there are interactions at the boundary that spoil the Gaussianity of the state, see for example \cite{ddb-21}). 
Hence, in our specific case, we start from the theory 
$\text{M-CFT} = \text{CFT}_1\otimes \dots \text{CFT}_{M}$ and we replicate it $n$ times, ending up with  $\text{M-CFT}^{\otimes n}$. 
Then, to compute ${\cal Z}_n$ we perform a diagonalisation in replica space and end up with the product of $n$ charged partition functions
which are given by Eq. \eqref{ZvsS} with the insertion of the appropriate flux, i.e. (with our normalisation ${\cal Z}_1=1$)
\be
\mathcal{Z}_1(\alpha) = \bra{S}e^{i\alpha Q_A}q^{L_0+\bar{L}_0}\ket{S}.
\label{eq:U(1)Defect}
\ee
Here the modular parameter
\be
q = \exp\l -\frac{2\pi^2}{\log\l L/\varepsilon \r} \r,
\ee
has been introduced for later convenience. Our goal then becomes the computation of
\begin{multline}
\mathcal{Z}_1(\alpha) = \prod_{k \in \mathbb{N}-1/2}\bra{0} \exp\l -i (S^\dagger)_{jj'}\overline{\Psi}^{j}_{k} {\Psi^\dagger}^{j'}_{k}+ (\Psi \leftrightarrow \Psi^\dagger)\r q^{L_0+\bar{L}_0}e^{i\alpha Q_A} \times \\
\exp\l i S_{jj'}{\Psi^\dagger}^j_{-k}\overline{\Psi}^{j'}_{-k} + (\Psi \leftrightarrow \Psi^\dagger)\r\ket{0},
\label{eq:U(1)Renyi}
\end{multline}
in the limit $q\rightarrow 1$, corresponding to $\frac{L}{\varepsilon}\rightarrow \infty$. For this purpose, we firstly decompose $S$, which is a unitary $M\times M$ matrix, in a block diagonal form
\be
S = \begin{pmatrix}S_{AA} & S_{AB} \\ S_{BA} & S_{BB}\end{pmatrix}.
\label{eq:S_block}
\ee
Here $A$ stands for the $M_A$ species belonging to the subsystem $A$, while $B$ refers to the remaining $M_B = M-M_A$ species. Further, we split the set of indices $j=1,\dots,M_A+M_B$, associated to all the species, in the following two sets
\be
a=1,\dots,M_A, \quad b=1,\dots,M_B
\ee
to shorthand the species of $A$ and $B$ respectively. In this way, the charge operator $Q_A$ is 
\be
Q_{A} = \sum_{k\in \mathbb{N}-1/2} \Psi^a_{-k}\Psi^a_{k} + \bar{\Psi}^a_{-k}\bar{\Psi}^a_{k} - (\Psi\rightarrow \Psi^\dagger),
\ee
where the summation over the index $a$ is understood.

We consider the contribution to the partition function \eqref{eq:U(1)Renyi} coming from the single Laurent mode $k$, which requires the evaluation of
\be
\bra{0} \exp\l -i (S^\dagger)_{jj'}\overline{\Psi}^{j}_{k} {\Psi^\dagger}^{j'}_{k}\r q^{L_0+\bar{L}_0}e^{i\alpha Q_A} \exp\l i S_{jj'}{\Psi^\dagger}^j_{-k}\overline{\Psi}^{j'}_{-k} \r\ket{0}.
\ee
The commutation relations among $q^{L_0+\bar{L}_0}e^{i\alpha Q_A}$ and the fermionic fields are
\be
\begin{split}
& q^{L_0+\bar{L}_0}e^{i\alpha Q_A} {\Psi^\dagger}^a_{-k} = e^{-i\alpha } q^{k}{\Psi^\dagger}^a_{-k}q^{L_0+\bar{L}_0}e^{i\alpha Q_A},\\
& q^{L_0+\bar{L}_0}e^{i\alpha Q_A} {\Psi^\dagger}^b_{-k} =  q^{k}{\Psi^\dagger}^b_{-k}q^{L_0+\bar{L}_0}e^{i\alpha Q_A},\\
& q^{L_0+\bar{L}_0}e^{i\alpha Q_A} \bar{\Psi}^a_{-k} =  e^{i\alpha}q^{k}\bar{\Psi}^a_{-k} q^{L_0+\bar{L}_0}e^{i\alpha Q_A},\\
& q^{L_0+\bar{L}_0}e^{i\alpha Q_A} \bar{\Psi}^b_{-k} =  q^{k}\bar{\Psi}^b_{-k} q^{L_0+\bar{L}_0}e^{i\alpha Q_A},
\end{split}
\label{eq:CommRel}
\ee
and they can be easily derived from the momentum/charge of the Laurent modes. 
Using $q^{L_0 + \bar{L}_0}e^{i\alpha Q_A}\ket{0} = \ket{0}$ and the commutation relations \eqref{eq:CommRel}, we get
\begin{multline}
\bra{0} \exp\l -i (S^\dagger)_{jj'}\overline{\Psi}^{j}_{k} {\Psi^\dagger}^{j'}_{k}\r q^{L_0+\bar{L}_0}e^{i\alpha Q_A} \exp\l i S_{jj'}{\Psi^\dagger}^j_{-k}\overline{\Psi}^{j'}_{-k} \r\ket{0} = \\
\bra{0} \exp\l -i (S^\dagger)_{jj'}\overline{\Psi}^{j}_{k} {\Psi^\dagger}^{j'}_{k}\r \times\\
\exp\l i q^{2k}S_{aa'}{\Psi^\dagger}^a_{-k}\overline{\Psi}^{a'}_{-k} + i q^{2k}S_{bb'}{\Psi^\dagger}^b_{-k}\overline{\Psi}^{b'}_{-k} + i e^{-i\alpha}q^{2k}S_{ab'}{\Psi^\dagger}^a_{-k}\overline{\Psi}^{b'}_{-k} + i e^{i\alpha}q^{2k}S_{ba'}{\Psi^\dagger}^b_{-k}\overline{\Psi}^{a'}_{-k} \r\ket{0}.
\end{multline}
The last expression can be evaluated (see Eq. \eqref{eq_BoundOverlap} in the Appendix), and we get
\begin{multline}
\bra{0} \exp\l -i (S^\dagger)_{jj'}\overline{\Psi}^{j}_{k} {\Psi^\dagger}^{j'}_{k}\r q^{L_0+\bar{L}_0}e^{i\alpha Q_A} \exp\l i S_{jj'}{\Psi^\dagger}^j_{-k}\overline{\Psi}^{j'}_{-k} \r\ket{0} = \\
\text{det}\l \begin{pmatrix} 1 & 0 \\ 0 & 1\end{pmatrix}+ q^{2k}\begin{pmatrix} S^\dagger_{AA} & S^\dagger_{BA} \\ S^\dagger_{AB}  & S^\dagger_{BB}  \end{pmatrix} \begin{pmatrix} S_{AA} & e^{-i\alpha}S_{AB} \\ e^{i\alpha}S_{BA}  & S_{BB}  \end{pmatrix}\r.
\end{multline}
Using the unitarity of $S$, $SS^\dagger = S^\dagger S =1$, one can show that (see Eq. \eqref{eq:rem_det} in the Appendix), 
\begin{multline}
\text{det}\l \begin{pmatrix} 1 & 0 \\ 0 & 1\end{pmatrix}+ q^{2k}\begin{pmatrix} S^\dagger_{AA} & S^\dagger_{BA} \\ S^\dagger_{AB}  & S^\dagger_{BB}  \end{pmatrix} \begin{pmatrix} S_{AA} & e^{-i\alpha}S_{AB} \\ e^{i\alpha}S_{BA}  & S_{BB}  \end{pmatrix}\r\propto \\
 \text{det}\l 1 + 2(S_{AA}^\dagger S_{AA} + (1-S_{AA}^\dagger S_{AA})\cos \alpha )q^{2k} +q^{4k} \r,
 \end{multline}
where the proportionality constant is an unimportant $\alpha$-independent prefactor (see the appendix). 
Putting all the pieces together and taking into account the contribution coming from exchanging $\Psi \leftrightarrow \Psi^\dagger$, 
we find the analytic expression of the $U(1)$ charged partition function in Eq. \eqref{eq:U(1)Renyi}
\be
\mathcal{Z}_1(\alpha) \propto \prod_{k \in \mathbb{N}-1/2} \text{det}\l 1 + 2(S_{AA}^\dagger S_{AA} + (1-S_{AA}^\dagger S_{AA})\cos \alpha )q^{2k} +q^{4k} \r^2.
\label{eq:U(1)1replica}
\ee
According to Eq. \eqref{eq:prod_flux}, the $n$-sheeted partition function $\mathcal{Z}_n$ can be written finally as
\begin{multline}
\mathcal{Z}_n = \prod^{\frac{n-1}{2}}_{p=-\frac{n-1}{2}}\mathcal{Z}_1(\alpha = 2\pi p/n) \propto \\ 
\prod^{\frac{n-1}{2}}_{p=-\frac{n-1}{2}}\prod_{k \in \mathbb{N}-1/2} \text{det}\l 1+2(S_{AA}^\dagger S_{AA} + (1-S_{AA}^\dagger S_{AA})\cos (2\pi p/n))q^{2k}+q^{4k}\r^2,
\label{eq:Zn_PartFun}
\end{multline}
which is the main result of this section, although not yet written in a very transparent form. 

From Eq. \eqref{eq:Zn_PartFun} it is clear that in the presence of several wires belonging to $A$, $M_A\geq 1$, there are $M_A$ factorised contributions depending on the eigenvalues of $1-S^\dagger_{AA}S_{AA}$ and coming from the presence of the determinant of a $M_A\times M_A$ matrix. In other words, if we define
\be
\mathcal{Z}_{n,\mathcal{T}_a} = \prod^{\frac{n-1}{2}}_{p=-\frac{n-1}{2}}\prod_{k \in \mathbb{N}-1/2} \l 1+2((1-\mathcal{T}_a) + \mathcal{T}_a\cos (2\pi p/n))q^{2k}+q^{4k}\r^2,
\label{eq:Zn_PartFun2}
\ee
as the contribution coming from the generic eigenvalue $\mathcal{T}_a \in \text{Spec}(1-S^\dagger_{AA}S_{AA})$, one has
\be
\mathcal{Z}_{n} = \prod_{a=1}^{M_A} \mathcal{Z}_{n,\mathcal{T}_a},
\ee
where ${\cal T}_a$ can be interpreted as generalised effective transmission probabilities.
Plugging this relation in the definition of the R\'enyi entropy in Eq. \eqref{eq:renyidef}, one gets
\be\label{eq:reduction}
S_{n}(A) = \sum_{a=1}^{M_A} S_{n,\mathcal{T}_a},
\ee
with
\be\label{eq:renyisum}
S_{n,\mathcal{T}_a} = \frac{1}{1-n}\log \frac{\mathcal{Z}_{n,\mathcal{T}_a}}{\mathcal{Z}^n_{1,\mathcal{T}_a}}
\ee
being the R\'enyi entropy associated to each $\mathcal{T}_a$.

For the sake of completeness, we provide the explicit result for the partition functions and for the entanglement entropies in the relevant limit $\frac{L}{\varepsilon}\to\infty$. 
Since the total entropy is just given by the sum of $M_A$ independent contributions with effective transmission ${\cal T}_a$ it is sufficient to write only one term. 
For convenience, we also define a parameter $\alpha'$, being a function of $\alpha$ and the effective transmission ${\cal T}_a$, satisfying 
\be
2\cos \alpha' = 2(1-{\cal T}_a + {\cal T}_a\cos \alpha ).
\label{eq:alphaprime}
\ee
The infinite product appearing in Eq. \eqref{eq:U(1)1replica} which gives the $U(1)$ partition function is explicitly evaluated in Appendix \eqref{sec_Jacobi},  obtaining
\be
\frac{\mathcal{Z}_1(\alpha)}{\mathcal{Z}_1(0)} = \l \frac{\theta_3\big( \frac{\alpha'}{2\pi},q \big)}{\theta_3(0,q)}\r^2.
\ee
In the limit $q\rightarrow 1$, the leading term of the partition function gives
\be
\log \frac{\mathcal{Z}_1(\alpha)}{\mathcal{Z}_1(0)} \simeq \frac{1}{\log q}\l \text{Li}_2(-e^{i\alpha'})+\text{Li}_2(-e^{-i\alpha'})-2\text{Li}_2(-1)\r = -\l \frac{\alpha'}{2\pi}\r^2\log \frac{L}{\varepsilon},
\label{eq:LogU(1)}
\ee
with $\alpha'$ given by \eqref{eq:alphaprime}. 
Summing over the $n$ values of the flux $\alpha$, one gets straightforwardly the $n$-th R\'enyi entropy plugging Eq. \eqref{eq:LogU(1)} into Eq. \eqref{eq:prod_flux}. 
After some long but simple algebra, the final result is
\be
S_{n,\mathcal{T}_a} =\left( \frac{2}{\pi^2(n-1)}
\sum_{p=1}^{\lfloor n/2 \rfloor} 
 \arcsin^2 \left(\sqrt{{\cal T}_a} \cos \frac{(2p-1)\pi}{2n}\right) \right) \log \frac{L}\varepsilon\,,
 \label{Cal-final}
\ee
which matches the one in Ref. \cite{cmv-12} where also the analytic analytical continuation to $n\rightarrow 1$ can be found and it is not repeated here.
We stress that the major advance in this section compared to the existing literature \cite{cmv-12,gm-17} has been to understand how the elements of $S_{AA}$
combine (via the eigenvalues of $(1-S^\dagger_{AA}S_{AA})$) to give the entanglement entropy of more than one wire, while previous studies focused on a single one. 
This result is also preparatory to the calculation of the negativity reported in the following section.

\section{CFT approach: Fermionic Negativity}
\label{sec:negativity}
In this section we apply the CFT formalism to the calculation of the negativity between two subsets of wires of a conformal junction. 
In particular, we consider here the partial time-reversal negativity (often just called fermionic negativity), 
which is a more suitable entanglement measure for fermionic systems in mixed 
states (see \cite{ssr-17,srrc-19,srgr-18,sr-19,sr-19a,ez-15,mbc-21,csg-19,mvdc-22}). 
We will proceed via the evaluation of the R\'enyi negativity for even $n=n_e$ and then we will study the replica limit $n_e \to 1$.

\subsection{R\'enyi negativities}

We consider the conformal junction of Fig. \ref{fig:star} with the subsystems $A,B$, and $C$ formed by three sets of wires.
We are interested in the negativity between $A$ and $B$. The definition in Eq. \eqref{negdef} is not well-suited to study the entanglement properties in
the context of fermionic systems (see \cite{ez-15,ssr-17,srrc-19}). 
To circumvent this issue, the partial time-reversal transformation of the reduced density matrix has been introduced \cite{ssr-17}
and here we use the same symbols $\rho_{AB}^{T_B}$ and $\cal E$, as for the standard partial transpose operation, having in mind that we refer always and only 
to the fermionic negativity.
The replica approach to the negativity \cite{cct-12,cct-13} starts from the computation of the moments of the partial transpose reduced density matrix that can be written
in terms of a ratio of partition functions as 
\begin{equation}
\mathrm{Tr} (\rho_{AB}^{T_B})^{n}=\frac{\hat{\mathcal{Z}}_n}{\hat{\mathcal{Z}}_1^n}.
\end{equation}
This path integral representation of the moments is similar to the one of the reduced density matrix in Eq. \eqref{ratioZ}, but here  $\mathcal{Z}_n$ 
is replaced by $\hat{\mathcal{Z}}_n$.
The latter is the partition function in the $n$-sheeted Riemann surface built in such a way to implement the partial time reversal transposition in the subsystem $B$ 
(see Refs. \cite{cct-12,cct-13} for more details on the partial transpose and \cite{ssr-17,srrc-19} for the fermion case). The negativity is finally obtained as \cite{cct-12,cct-13}
\begin{equation}
    \mathcal{E}=\lim_{n_e\to 1}\mathrm{Tr}(\rho_{AB}^{T_B})^{n_e},
\end{equation}
i.e. by taking the analytic continuation from the even sequence of replicas, $n=n_e$.
$\hat{\mathcal{Z}}_n$ can be further factorised using the replica diagonalisation, such that it becomes the product of $n$ single-replica $U (1)$ charged partition functions, similarly to what has been done for $\mathcal{Z}_n$ in the previous section, but with some differences. 
Let us focus on even $n=n_e$, which is the only necessary object to compute the negativity (instead for the negativity spectrum, i.e. the spectrum of $\rho_{AB}^{T_B}$ 
also odd values of $n$ matters \cite{srrc-19,rac-16b}, as well as for other entanglement witnesses \cite{ekh-20,ncv-21}). 
The needed charged partition $\hat{\mathcal{Z}}_1(\alpha)$ has twisting phases equal to $e^{i\alpha}$ in $A$ and $e^{i(\pi -\alpha)}$ in $B$, i.e. it reads \cite{ssr-17,sr-19a}
\begin{equation}
\hat{\mathcal{Z}}_1(\alpha)=\braket{0|e^{i\alpha Q_A}e^{-i(\alpha-\pi)Q_B}|0}.
\end{equation}
The operator $e^{i\alpha Q_A}$ implements the $U(1)$ symmetry restricted to $A$, while $e^{-i(\alpha-\pi)Q_B}$ inverts the flux ($\alpha \to -\alpha$) and it introduces an additional phases $-1$ along $B$, which is the combined net effect of the partial transpose operation (or, equivalently, partial time reversal) on fermionic systems. 
The final result of this approach is that the $n_e$-th R\'enyi negativity can be computed as
\be
\mathcal{E}_{n_e} \equiv \log \text{Tr}\l |\rho^{T_B}_{AB}|^{n_e}\r =\log \text{Tr}\l (\rho^{T_B}_{AB})^{n_e}\r 
= \sum^{\frac{n_e-1}{2}}_{p=-\frac{n_e-1}{2}} \log \frac{\hat{\mathcal{Z}}_1(\alpha=2\pi p/n_{e})}{\hat{\mathcal{Z}}_1(0)}.
\label{eq:Renyi_Neg_Sum}
\ee
In the presence of boundaries,  the $U(1)$ charged  partition function straightforwardly becomes
\begin{multline}
\hat{\mathcal{Z}}_1(\alpha) = \bra{S}q^{L_0+\bar{L}_0}e^{i\alpha Q_A -i(\alpha-\pi)Q_B}\ket{S} = \\
\prod_{k \in \mathbb{N}-1/2} \bra{0} \exp\l -i (S^\dagger)_{jj'}\overline{\Psi}^{j}_{k} {\Psi^\dagger}^{j'}_{k} + (\Psi \leftrightarrow \Psi^\dagger)\r q^{L_0+\bar{L}_0}\times\\
e^{i\alpha Q_A-i(\alpha-\pi) Q_B} \exp\l i S_{jj'}{\Psi^\dagger}^j_{-k}\overline{\Psi}^{j'}_{-k} + (\Psi \leftrightarrow \Psi^\dagger) \r\ket{0},
\label{eq:U(1)negativity}
\end{multline}

Eq. \eqref{eq:Renyi_Neg_Sum} with \eqref{eq:U(1)negativity} holds for a generic tripartite fermionic system with no assumption.
To proceed for the calculation of the wire junction we restrict to the following specific situation: 
\begin{itemize}
    \item $M_A,M_B,M_C=1$, so that the total number of wires is $M=3$.
    \item $S$ is not only unitary but also Hermitian, which means that $S^2=1$ and its eigenvalues can be just $\pm 1$.
    For some physical systems (including the Schrodinger junction in the next section), the hermiticity of the S matrix is a necessary condition for physical consistency.
    Hence this is not at all a very restrictive assumption.  
\end{itemize}
With these working assumptions it is possible to obtain nice analytic results in a rather compact form. 
More general expressions (e.g. for more wires) can also be obtained, but at the price of more cumbersome computations and less intelligible final results 
without any major physical insight.

It is clear from Eq. \eqref{eq:U(1)negativity} that for $M=3$ the key object to be evaluated is
\begin{multline}\label{eq:det}
\bra{0} \exp\l -i S_{jj'}\overline{\Psi}^{j}_{k} {\Psi^\dagger}^{j'}_{k}\r q^{L_0+\bar{L}_0}e^{i\alpha Q_A-i(\alpha -\pi)Q_B} \exp\l i S_{jj'}{\Psi^\dagger}^j_{-k}\overline{\Psi}^{j'}_{-k} \r\ket{0} =\\
\text{det}\l  1+  q^{2k}S \begin{pmatrix} S_{AA} & -e^{-i2\alpha}S_{AB} & e^{-i\alpha}S_{AC} \\ -e^{i2\alpha}S_{BA}  & S_{BB} & -e^{i\alpha}S_{BC}\\
e^{i\alpha}S_{CA} & -e^{-i\alpha}S_{CB} & S_{CC}
\end{pmatrix}\r,
\end{multline}
where we used the commutation relations between $q^{L_0+\bar{L}_0}e^{i\alpha Q_A-i(\alpha -\pi)Q_B}$ and the fields (see Eq. \eqref{eq:CommRel}), and the formula \eqref{eq:rem_det}, which provides the vacuum expectation value. The determinant of the $3\times 3$ matrix appearing in \eqref{eq:det} can be evaluated directly, but it is useful to discuss first the constraints due to the unitarity of $S$. We define the matrix $\mathcal{O}$ as 
\be
\mathcal{O} = S \begin{pmatrix} S_{AA} & -e^{-i2\alpha}S_{AB} & e^{-i\alpha}S_{AC} \\ -e^{i2\alpha}S_{BA}  & S_{BB} & -e^{i\alpha}S_{BC}\\
e^{i\alpha}S_{CA} & -e^{-i\alpha}S_{CB} & S_{CC} \end{pmatrix},
\ee
and we verify the following properties:
\begin{itemize}
    \item $\mathcal{O}$ is unitary ($\mathcal{O}\mathcal{O}^\dagger=1$) and its eigenvalues are phases;
    \item $\text{det}(\mathcal{O})=1$ and the product of the eigenvalues is $1$;
    \item $\mathcal{O} = S\mathcal{O}^\dagger S$ and the spectrum of $\mathcal{O}$ is thus invariant under complex conjugation, a feature that relies on our hermiticity assumption $S=S^\dagger$.
\end{itemize}
These properties imply that the spectrum of $\mathcal{O}$ has to take this form
\be
\text{Spec}\l \mathcal{O}\r = \{1,e^{i\hat{\alpha}'},e^{-i\hat{\alpha}'}\},
\label{eq:SpecO}
\ee
with $\hat{\alpha}'$ a real parameter depending on $S$ and $\alpha$, defined by the following property
\be
2\cos\hat{\alpha}' +1 = \text{Tr}(\mathcal{O}),
\ee
which is a consequence of Eq. \eqref{eq:SpecO}. The determinant $\text{det}(1+q^{2k}\mathcal{O})$ can be thus computed taking the product over the eigenvalues of $\mathcal{O}$ as follows
\be\label{eq:det1}
\text{det}\l 1+q^{2k}\mathcal{O}\r = (1+q^{2k})(1+q^{2k}e^{i\hat{\alpha}'})(1+q^{2k}e^{-i\hat{\alpha}'}) = (1+q^{2k})(1+q^{2k}(\text{Tr}(\mathcal{O})-1)+q^{4k}).
\ee
Evaluating $\text{Tr}(\mathcal{O})$ and using again the unitarity of $S$, we can rewrite Eq. \eqref{eq:det1} as
\be
\text{det}\l 1+q^{2k}\mathcal{O}\r = (1+q^{2k})(1+2\cos \hat{\alpha}'q^{2k}+q^{4k}),
\ee
and the explicit expression of $\hat{\alpha}'$ as a function of the $S$ matrix is
\be
2\cos \hat{\alpha}' = -1+ S^2_{AA} + S^2_{BB} + S_{CC}^2+ (-1-S^2_{CC} +S^2_{BB}+S^2_{AA})\cos(2\alpha) + 2(S_{BB}^2-S_{AA}^2)\cos \alpha.
\label{eq:alphaHat}
\ee
Notice that $\hat{\alpha}'$ does only depend on the diagonal entries of the matrix $S$ and on the flux $\alpha$.\\ Putting all the pieces together, we express the partition function $\hat{\mathcal{Z}}_1(\alpha)$ as
\be
\hat{\mathcal{Z}}_1(\alpha) = \prod_{k \in \mathbb{N}-1/2} (1+q^{2k})^2(1+2\cos \hat{\alpha}'q^{2k}+q^{4k})^2.
\ee
We find the same formal structure of the partition function which appeared for the R\'enyi entropies in Eq. \eqref{eq:U(1)1replica}, up to the replacement $\alpha'\rightarrow \hat{\alpha}'$. Analogously to Eq. \eqref{eq:LogU(1)}, for $L/\epsilon \gg1$, we have
\be
\log \frac{\hat{\mathcal{Z}}_1(\alpha)}{\hat{\mathcal{Z}}_1(0)} \simeq  -\l \frac{\hat{\alpha}'}{2\pi}\r^2\log \frac{L}{\varepsilon},
\label{eq:LogU(1)Hat}
\ee
with $\hat{\alpha}'$ given by Eq. \eqref{eq:alphaHat}, which is the main result of this section. Indeed, by plugging this result into 
Eq. \eqref{eq:Renyi_Neg_Sum}, we obtain the R\'enyi negativities
\begin{multline}\label{eq:cal-final}
{\cal E}_{n_e}=-\Big( \frac{1}{4\pi^2}
\sum_{p=-(n_e-1)/2}^{(n_e-1)/2} 
 \arccos^2 \left(S_{CC}^2+
 (-1-S^2_{CC} +S^2_{BB}+S^2_{AA})\cos(2\pi p/n_e)^2+ \right. \\  \left.+ (S_{BB}^2-S_{AA}^2)\cos(2\pi p/n_e) \right)\Big) \log \frac{L}\varepsilon\,.
\end{multline}
We conclude this subsection by providing few simple consistency checks for $\hat{\alpha}'$ in some limits. 
If the wire $C$ is decoupled from the other two, then $S_{CC}^2 = 1$ and the transmission probability between $A$ and $B$ is $1-S^2_{AA} = 1-S^2_{BB}$. In that case
\be
2\cos \hat{\alpha}'= 2S^2_{AA}-2(1-S_{AA}^2)\cos(2\alpha).
\ee
This value of $\hat{\alpha}'$ is the same one would obtain for $\alpha'$ in the $U(1)$ partition function of $A$ and $B$ in the presence of a flux $-e^{i2\alpha}$ inserted along $A$. The reason is that in this limit the system becomes invariant under the $U(1)$ symmetry generated by $Q_A+Q_B$ and, thanks to
\be
e^{i\alpha Q_{A}-i(\alpha-\pi)Q_B} = e^{-i(\alpha-\pi)(Q_B+Q_A)}e^{i(2\alpha-\pi)Q_A},
\ee
one recognises an equivalence with the insertion of $e^{i(2\alpha-\pi)Q_A}$.

Finally, we consider $\alpha = \pi/2$ because it corresponds to the evaluation of the $2$-R\'enyi negativity 
(indeed, from Eq. \eqref{eq:Renyi_Neg_Sum}, ${\cal E}_2=\log (\hat {\cal Z}_1(\pi/2)\hat {\cal Z}_1(-\pi/2)/\hat{\cal Z}_1^2(0))= 2 \log \hat {\cal Z}_1(\pi/2)/\hat{\cal Z}_1(0)$). 
In this case
\be
2\cos \hat{\alpha}' = 2S^2_{CC},
\ee
and there is no explicit dependence on $S_{AA},S_{BB}$. Now this parameter is the same $\alpha'$ one would get in the presence of a flux $e^{i\pi/2}$ along $C$ only. 
In this way, we reproduced the general identity \cite{cct-12}
\be
\text{Tr}\big( \big(\rho_{AB}^{T_B}\big)^2\big) = \text{Tr}\big( \big(\rho_{AB}\big)^2\big),
\ee
which is the well known relation between the $2$-R\'enyi negativity and the $2$-R\'enyi entropy.

\subsection{Analytic continuation}

The representation of $\mathcal{E}_{n_e}$ as a sum, appearing in Eq. \eqref{eq:cal-final}, gives an expression valid only when $n_e$ is an even natural number.
To proceed to the calculation of the negativity, we should provide its analytic continuation for $n_e$ being a generic number in the complex plane. 
To this goal, the strategy we device is the following:
\begin{itemize}
    \item Using the identities of Appendix \ref{sec_Jacobi},
    the $U(1)$ partition function $\hat{\mathcal{Z}}_1(\alpha)$ in Eq. \eqref{eq:LogU(1)Hat} can be expressed through an integral representation in the limit $q \to 1$ 
    \begin{multline}
    \log \frac{\hat{\mathcal{Z}}_1(\alpha)}{\hat{\mathcal{Z}}_1(0)}=\sum_{k \in \mathbb{N}-1/2}2\log [ (1+q^{2k})^{-2}(1+2\cos \alpha'q^{2k}+q^{4k})]\\ \simeq
    -\frac{1}{\log q}\int_0^{\infty}\frac{dt}{t} [\log (1+2\cos \alpha't+t^{2})-2\log(1+t^{2})];
    \end{multline}
    \item The sum over the value of fluxes \eqref{eq:cal-final} can now be performed inside the integral. Through some simple trigonometric identities, this leads to an analytic continuation of the integrand.
    \item The final result is an integral, which represents our analytic continuation.
\end{itemize}
As aforementioned, using the trigonometric identities studied in the appendix \ref{sec:trig}, we find
\be
(1+2\cos \hat{\alpha}' t + t^{2})=\left(\cos(\alpha)+\frac{b-\sqrt{b^2-4 ac}}{2c}\right)\left(c\cos(\alpha)+\frac{b+\sqrt{b^2-4 ac}}{2}\right),
\ee
with $a,b,c$ being the following functions of the $S$ matrix and $t$
\be
a=1+2 S_{CC}^2t+t^{2},\qquad b=2(S_{BB}^2-S_{AA}^2)t,\qquad c=2(S^2_{AA} + S^2_{BB} - S_{CC}^2-1)t.
\ee
Using the integral representation for the product over the $k$ modes in the limit $q \to 1$ reported in \ref{sec_Jacobi}, we get
\begin{multline}\label{eq:negativity}
\mathcal{E}_{n_e} = \sum^{\frac{n_e-1}{2}}_{p=-\frac{n_e-1}{2}}\log \frac{\hat{\mathcal{Z}}_1(\alpha)}{\hat{\mathcal{Z}}_1(0)}= \frac{\log (L/\varepsilon)}{2\pi^2}\int_0^1 \frac{dt}{t}\\ \times \sum^{\frac{n_e-1}{2}}_{p=-\frac{n_e-1}{2}}\left[ \log \left(\cos(2\pi p/n_e)+\frac{b-\sqrt{b^2-4 ac}}{2c}\right)\left(c\cos(2\pi p/n_e)+\frac{b+\sqrt{b^2-4 ac}}{2}\right)-2\log(1+t)\right]\\
=\frac{\log (L/\varepsilon)}{\pi^2}\int_0^1 \frac{dt}{t}\\ \times \log \frac{ ((x_1 - \sqrt{-c^2 + x_1^2})^{n_e/2} + (
   x_1 + \sqrt{-c^2 + x_1^2})^{n_e/2}) ((x_2 - \sqrt{-1 + x_2^2})^{n_e/2} + (
   x_2 + \sqrt{-1 + x_2^2})^{n_e/2})}{2^{n_e}(1+t)^{n_e}}
\end{multline}
where
\be
x_1=\frac{b+\sqrt{b^2-4ac}}{2}, \qquad x_2=\frac{b-\sqrt{b^2-4ac}}{2c}.
\ee
Eq. \eqref{eq:negativity} is the desired analytic continuation. 
At this point, the negativity is simply obtained by taking $n_e \to 1$:
\begin{multline}
{\cal E}= \frac{\log (L/\varepsilon)}{\pi^2}\int_0^1 \frac{dt}{t}   \\
\log \frac{ ((x_1 - \sqrt{-c^2 + x_1^2})^{1/2} + (
   x_1 + \sqrt{-c^2 + x_1^2})^{1/2}) ((x_2 - \sqrt{-1 + x_2^2})^{1/2} + (
   x_2 + \sqrt{-1 + x_2^2})^{1/2})}{(1+t)}.
\end{multline}

 Let us conclude this section by providing some useful cross-checks of our result. In the limit in which $S^2_{CC}=1$, the wire $C$ decouples and one recovers the result for two wires, which is given by the R\'enyi entropy with $n=1/2$ ($A$ and $B$ now form a pure state) \cite{ge-20}. 
  In this case, Eq. \eqref{eq:negativity} simplifies as
\begin{equation}
    \mathcal{E}=\frac{1}{\pi^2}\int_0^1\frac{dt}{t}\log \bigg(1+\frac{2\sqrt{t}\sqrt{1-S^2_{AA}}}{1+t}\bigg)=\frac{1}{\pi^2}\arcsin\l\sqrt{1-S^2_{AA}}\r\l \pi-\arcsin \sqrt{1-S^2_{AA}}\r.
\end{equation}
Performing the change of variables $t=e^{2x}$, one can recover the result (in the integral form) for the R\'enyi 1/2 obtained in \cite{cmv-12} or, by solving the integral, the result for the negativity between two CFTs in \cite{ge-20}, with $s=\sqrt{1-S^2_{AA}}$.

\section{Schroedinger junction}
\label{sec:schroedinger}

In this section, we describe a fermion gas on a star graph modelling a junction made up of $M$ wires of length $L$, joined together through a single defect. 
We introduce a slightly different framework (compared to the existing ones in the literature) which allows us to efficiently perform exact numerical computations 
also for the negativity. 
The CFT predictions of the previous sections are checked against these exact numerical results.

\subsection{Correlation functions}
Let us consider a star graph like the one in Fig. \ref{fig:star} where now on each wire there is a gas made of $N$ spinless fermions. 
The $M$ wires are decoupled everywhere but in the vertex of the graph and their mixing is described by a non-trivial scattering matrix. 
We consider the ground state of such system with $N$ particles.
The same system has been studied in Ref. \cite{cmv-12} by the overlap matrix approach \cite{cmv-11,cmv-11b} which is the starting point of our analysis. 
Each point of the junction is parametrised by a pair
\be
(x,j), \quad x \in [0,L], \quad j = 1, \dots,M,
\ee
where $j$ is the index identifying the wire and $x>0$ the spatial coordinate along the wire. 
The bulk hamiltonian of the system is
\be
H = \sum^{M}_{j=1} \int^L_{0} dx\frac{1}{2}\l \partial_x \Psi^\dagger_j(x)\r\l \partial_x \Psi_j(x)\r,
\ee
with $\Psi_j,\Psi_j^\dagger$ being the fermionic fields associated to the $j$-th wire (also called Schroedinger field, from which the name Schroedinger junction). 
We consider a scattering matrix
\be
S_{ij}, \quad i,j = 1,\dots, M,
\ee
describing the defect at $x=0$, which has to be hermitian and unitary \cite{bm-06,cmv-12}
\be
S=S^\dagger, \quad SS^\dagger =1.
\ee
The most general boundary condition along the junction is
\begin{equation}
\lambda(1-S)\Psi(0)-i(1+S)\partial_x\Psi(0)=0,
\end{equation}
where $\Psi=\{\Psi_j\}_{j=1,\dots,M}$,  $\lambda$ is an arbitrary real parameter with the dimension of mass.
To fully specify the problem, we also need to impose boundary conditions at the external edges of each wire that generically take the form
\be
(\partial_x \Psi_i)(L)=\mu_i\Psi_i(L),
\ee
where $\mu_i$ are again real parameters with the dimension of mass. In order to simplify the treatment, it is possible to diagonalise $S$ via a unitary transformation $\mathcal{U}$ and its eigenvalues are just $\pm 1$. It is custom \cite{bm-06,bms-07} to introduce a set of unphysical fields $\{\varphi_j(x)\}$
\be
\Psi_i(x)= \sum_{j=1}^M\mathcal{U}_{ij}\varphi_j(x),
\ee
so that in terms of these $\varphi_j(x)$, the boundary conditions decouple as
\be
\begin{split}
&\partial_x \varphi_i(0)=\eta_i\varphi_i(0),\\
&\partial_x \varphi_i(L)=\mu'_i\varphi_i(L),
\end{split}
\ee
where the $\mu'_i$'s are linear function of the $\mu_i$'s whose form is irrelevant. 
For the junction to be scale-invariant, we require each of the dimensionful parameters $\eta_i$ and $\mu_i$ to be either $0$ or $\infty$. 
The choice corresponds to either Neumann ($\partial_x\varphi_i=0$) or Dirichlet ($\varphi_i =0$) boundary conditions at $x=0$ and $x=L$. 
We impose Dirichlet boundary conditions ($\mu_i=\infty$) at $x=L$ for all wires. 
Conversely, the values of $\eta_i$ being $0$ or $\infty$ depends on the diagonalisation of the $S$ matrix, see \cite{bm-06,bms-07,cmv-12}. 
Hence, the unphysical fields $\varphi_j(x)$ have Dirichlet bc's at $x=L$ and either (Neumann or Dirichlet) at $x=0$, so that it is natural to 
use the short-hand notation
\be
ND \quad \text{Neumann-Dirichlet}, \quad DD \quad \text{Dirichlet-Dirichlet},
\ee
to refer to the two possibbilities.
For these two possible boundary conditions, the single-particle wavefunctions are
\be
\begin{split}
&\phi^{DD}(n,x) = \sqrt{\frac{2}{L}}\sin \frac{n\pi x}{L}, \quad n=1,\dots\\
&\phi^{ND}(n,x) = \sqrt{\frac{2}{L}}\cos \frac{\l n-1/2\r\pi x}{L}, \quad n=1,\dots.
\end{split}
\ee
We work in the ground state with fixed particle number $N$ for each wire, so that the correlation function is
\be
\la \varphi_i(x)\varphi_j(y)\ra = \delta_{ij} \times \begin{cases} C_{DD}(x,y), \quad DD \text{ bc's}\\C_{ND}(x,y), \quad ND \text{ bc's}, \end{cases}
\ee
with
\be
\begin{split}
&C_{DD}(x,y) = \sum^{N}_{n=1} \phi^{DD}(n,x)\overline{\phi^{DD}(n,y)} = \frac{\sin \frac{N+1/2}{ L}\pi(x-y)}{2L\sin \frac{\pi(x-y)}{2L}} -(y\rightarrow -y)\\
&C_{ND}(x,y) =  \sum^{N}_{n=1} \phi^{ND}(n,x)\overline{\phi^{ND}(n,y)} = \frac{\sin \frac{N}{ L}\pi(x-y)}{2L\sin \frac{\pi(x-y)}{2L}} + (y\rightarrow -y).
\end{split}
\ee
Going back to the physical fields $\{\Psi_j\}_j$, linear algebra straightforwardly gives
\be
C_{ij}(x,y) \equiv \la \Psi_j^\dagger(x)\Psi_i(y) \ra = \l \frac{1+S}{2}\r_{ij} C_{ND}(x,y) + \l \frac{1-S}{2}\r_{ij}C_{DD}(x,y).
\label{eq_Kernel}
\ee
The matrices $\frac{1 \pm S}{2}$ are the projectors over the eigenspaces of $S$ with eigenvalues $\pm 1$ respectively. 
In fact, given any eigenvector $v_\pm$ of $S$ satisfying $S v_{\pm} = \pm v_{\pm},$,  by inspection it holds
\be
\frac{1\pm S}{2} v_{\pm} = v_{\pm}, \qquad \frac{1\mp S}{2} v_{\pm} = 0.
\ee

\subsection{Finite-dimensional representation of the correlation function}

The correlation functions \eqref{eq_Kernel} are continuous kernel of the spatial variables. 
While it is possible to work directly with such kernels (as done, e.g., in Refs. \cite{clm-14,v-12}), it is more convenient to work with a finite-dimensional representation of such correlation.
In this subsection we derive a representation which is particularly useful for numerical applications and it is equivalent to the 
overlap matrix approach \cite{cmv-11}. The main result can be read in Eq. \eqref{eq_MatrixC}. 

Hereafter, we set $L=1$ without loss of generality. 
We start noticing that $C_{ij}(x,y)$ can be thought as an operator acting on the Hilbert space $\mathbb{C}^M \otimes L^2([0,1])$,
with $\mathbb{C}^M$ representing the space of the wires  and $L^2([0,1])$ being the one of wave-functions on $[0,1]$. 
Although $L^2([0,1])$ is an infinite-dimensional Hilbert space, both $C_{ND}$ and $C_{DD}$ are projectors acting non-trivially only in a finite-dimension subspace $\mathcal{H}_0$.
We can choose the following basis for $\mathcal{H}_0$
\be
e_n \equiv \begin{cases} \phi^{DD}(n,x), \quad 1\leq n \leq N \\ \phi^{ND}(n-N,x), \quad 1+N\leq n \leq 2N,
\label{eq_en}
\end{cases}
\ee
which is not orthonormal. Indeed, since the single-particle eigenfunctions are normalised in both ND and DD sectors, one has
\be
\la e_n,e_{n'} \ra = \int dx \  \overline{\phi^{DD}(n,x)}\phi^{DD}(n',x) = \delta_{n,n'}, \quad n,n' = 1,\dots,N,
\ee
and similarly if $n,n' = 1+N,\dots,2N$. Instead, for $n\leq N$ and $n'\geq N+1$, their scalar product $Q_{n,n'}$ is 
\be
Q_{n,n'} \equiv \la e_n,e_{n'} \ra  = 2 \int^1_0 dx \sin\l n\pi x\r\cos\l \l n' -\frac{1}{2}\r\pi x\r = \frac{2 n}{\pi\l n^2-(n'-1/2)^2\r}.
\label{eq_MatrixQ}
\ee
Since the basis is not orthonormal, we have to be careful to correctly give a matrix representations of $C_{ND}(x,y),C_{DD}(x,y)$, and thus of $C_{ij}(x,y)$. We introduce the dual space $\mathcal{H}^*_0$, as the space of linear functional on $\mathcal{H}_0$, and the dual basis
\be
\{e^*_n\}_{n=1,\dots,2N},
\ee
defined by
\be
e^*_n(e_{n'}) = \delta_{nn'}.
\ee
To avoid confusion we use another symbol for the bra associated to $e_n$, which is denoted by
\be
e^\dagger_n \in \mathcal{H}^*_0,
\ee
and it is defined by
\be
e^\dagger_n(v) \equiv \la e_n,v \ra, \quad \forall v \in \mathcal{H}_0.
\ee
We stress that $e^*_n\neq e^\dagger_n$, since
\be
e^\dagger_n(e_{n'}) = Q_{n,n'}, \quad  e^*_n(e_{n'}) = \delta_{nn'}  ,\quad n\leq N, n'\geq N+1,
\ee
which is a consequence of the non-orthonormality of the basis. The projectors $C_{ND}$ and $C_{DD}$, seen as operators of $\mathcal{H}_0$ and belonging to the space
\be
\text{End}\l \mathcal{H}_0 \r \simeq \mathcal{H}_0\otimes  \mathcal{H}^*_0,
\ee
have the following expression
\be
C_{DD} = \sum^{N}_{n=1} e_n\otimes e^\dagger_n, \quad C_{ND} = \sum^{2N}_{n=N+1} e_n\otimes e^\dagger_n.
\ee
In the basis $\{e_n\}^{2N}_{n=1}$ of $\mathcal{H}_0$, they are represented by the following matrix elements
\be
\begin{split}
&\l C_{DD} \r_{n,n'} \equiv e^*_n(C_{DD} \ e_{n'}) = \sum^N_{m=1} e^*_n(e_m) e_m^\dagger(e_{n'}) = \sum^N_{m=1} \delta_{n,m} \la e_m,e_{n'} \ra\\
&\l C_{ND} \r_{n,n'} \equiv e^*_n(C_{ND} \ e_{n'}) = \sum^{2N}_{m=N+1} e^*_n(e_m) e_m^\dagger(e_{n'}) = \sum^{2N}_{m=N+1} \delta_{n,m} \la e_m,e_{n'} \ra,
\end{split}
\ee
which means that their associated $2N\times 2N$ matrices are
\be
C_{DD} = \begin{pmatrix} 1 & Q\\ 0 & 0\end{pmatrix}, \quad C_{ND}= \begin{pmatrix} 0 & 0\\ Q^\dagger & 1\end{pmatrix}.
\ee
In conclusion, we represented the correlation function $C_{ij}(x,y)$ in Eq. \eqref{eq_Kernel} as a $(2MN,2MN)$ matrix acting on 
\be
\mathbb{C}^{M}\otimes \mathcal{H}_0 \simeq \mathbb{C}^{M}\otimes \mathbb{C}^{2N},
\ee
as follows
\be
C =\frac{1-S}{2} \otimes \begin{pmatrix} 1 & Q\\ 0 & 0\end{pmatrix} + \frac{1+S}{2} \otimes \begin{pmatrix} 0 & 0\\ Q^\dagger & 1\end{pmatrix},
\label{eq_MatrixC}
\ee
with $Q$ being a $N\times N$ matrix defined by Eq. \eqref{eq_MatrixQ}.

\subsection{The R\'enyi entropy between two arbitrary sets of wires}\label{sub:overlap}

A useful auxiliary quantity for the computation of the entanglement entropy and negativity is the matrix $\Gamma = 1-2C$ (sometimes referred to as covariance matrix).
Using Eq. \eqref{eq_Kernel} and the finite-dimensional representation of the correlation matrix in the previous section, we can 
express it as
\be
1-\Gamma^2 = 4C(1-C) = (1-S^2)\otimes (C_{ND}-C_{DD})^2.
\label{eq:1-Gamma^2}
\ee
When $\Gamma$ refers to the entire system, given that $S^2=1$, it is zero in the ``wire space''.
However, Eq. \eqref{eq:1-Gamma^2} has a very convenient form for the restriction of $\Gamma$ to a subsystem $A$ made of $M_A$  wires because of the tensor product 
structure in internal and spatial coordinates. It is in fact enough to replace $S\rightarrow S_{AA}$, i.e. the projected $S$-matrix to obtain $\Gamma_{AA}$, the 
covariance matrix restricted to the subsystem of interest:
\be
1-\Gamma_{AA}^2 = (1-S_{AA}^2)\otimes (C_{ND}-C_{DD})^2,
\ee
with $S_{AA}$ being a $M_A \times M_A$ matrix.
In general $1-S_{AA}^2 \geq 0$, and so $1-\Gamma_{AA}^2$ is positive semidefinite.
We recall that the matrix representation for $(C_{ND}-C_{DD})^2$ is
\be
(C_{ND}-C_{DD})^2 = \begin{pmatrix} 1-QQ^\dagger & 0 \\ 0 & 1-Q^\dagger Q\end{pmatrix}.
\ee
Since $1-QQ^\dagger$ and $1-Q^\dagger Q$ have the same spectrum, the spectrum of $(C_{ND}-C_{DD})^2$ is obtained by two copies of the spectrum of $1-QQ^\dagger$.

To start, let us recover the results \cite{cmv-12} for the case when $A$ is a single wire. 
First, taking $M=2$ and the completely transmissive $S$-matrix 
\be
S = \begin{pmatrix} 0 & 1 \\ 1 & 0\end{pmatrix},
\ee
the correlation function of a single wire $A$ satisfies
\be
1-\Gamma_{AA}^2 = (C_{ND}-C_{DD})^2.
\ee
The spectral properties of $1-\Gamma_{AA}^2$ are then the one of a gas on the line $[-L,L]$ bipartite as $A = [0,L]$ and $B = [-L,0]$. 
For a generic $S$-matrix of two wires \cite{bddo-02,bm-06}
\be
S = \begin{pmatrix} \sqrt{1-|s|^2} & se^{i\phi} \\ \bar{s}e^{-i\phi} & -\sqrt{1-|s|^2}\end{pmatrix},
\label{eq:SMatrixRenyi}
\ee
$\Gamma_{AA}$ is 
\be
1-\Gamma_{AA}^2 = |s|^2 (C_{ND}-C_{DD})^2,
\ee
and so the eigenvalues of $1-\Gamma_{AA}^2$ are just rescaled by a factor $|s|^2$, in agreement with what known \cite{cmv-12} from the overlap matrix.
In the case of $A$ being one of the $M$ wires in a junction, in the above equation is enough to replace $s$ with the transmission coefficient of $A$. 

Once the covariance matrix $\Gamma_{AA}$ is known, the entanglement R\'enyi entropy is \cite{ep-09}
\be
S_{n}(A) = \frac{1}{1-n}\text{Tr}\log \l \l \frac{1+\Gamma_{AA}}{2}\r^{n} + \l \frac{1-\Gamma_{AA}}{2}\r^{n} \r  = \sum_{\mathcal{T}_a}S_{n,\mathcal{T}_a},
\ee
Using the basic property about the spectrum of a tensor product 
\be
\text{Spec}(X\otimes Y) = \{ x_iy_j \}_{ij}, \quad x_{i} \in \text{Spec}(X), \quad \quad y_{j} \in \text{Spec}(Y), 
\ee
we get that the R\'enyi entropy between $A$ and the complementary wires $B$ is
\be
S_{n}(A) = \sum_{\mathcal{T}_a}S_{n,\mathcal{T}_a},
\ee
where the $\mathcal{T}_a$'s are the eigenvalues of $1-S_{AA}^2$, which play the role of a transmission probability, 
and $S_{n,\mathcal{T}_a}$ is the R\'enyi entropy of a single wire with transmission probability $\mathcal{T}_a$. 
This analytic results for the microscopic model perfectly match Eq. \eqref{Cal-final} in CFT.  
We also notice that the relation $S_n(A) = S_n(B)$ comes from the fact that $1-S^2_{AA}$ and $1-S_{BB}^2$ have the same non-zero spectrum 
(i.e., the same eigenvalues up to the vanishing ones). 

\begin{figure}[t]
\centering
	\includegraphics[width=0.6\linewidth]{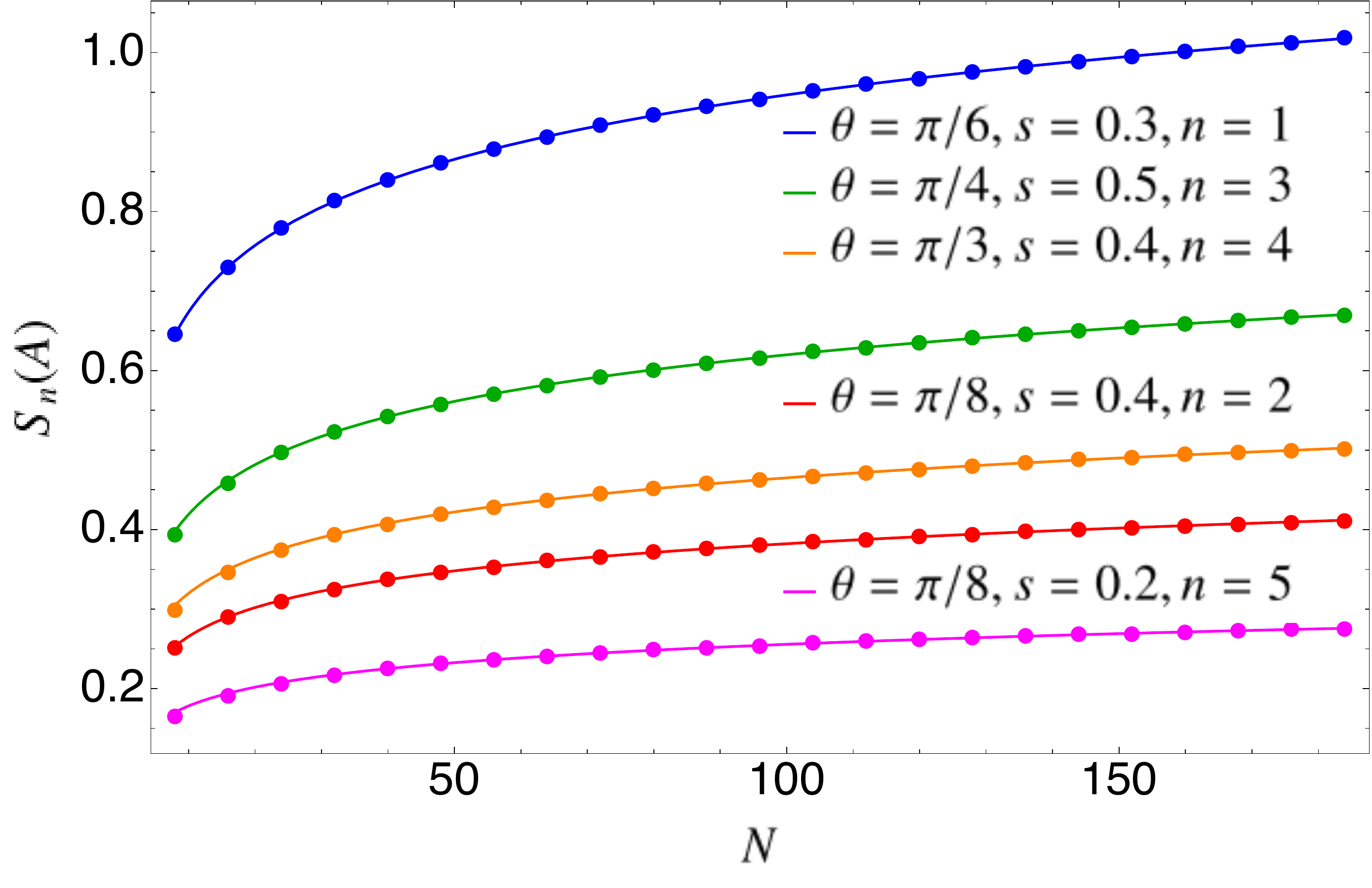}
    \caption{The R\'enyi entropies $S_n(A)$ in a four-wire junction where $A$ is made up of two wires.
We choose different values of $s, \theta,n$ and we plot it as a function of the number of particles $N$. The lines show the curve $C_n (s, \theta) \log N + b_0 + b_1N^{-1/n}$ where the coefficients $b_i$ are fitted using the data for $N \geq 80$. The coefficients $C_n (s, \theta)$ are obtained by summing over the single-wire results, as explained in Eq. \eqref{Cal-final}}
    \label{fig:n0}
\end{figure}

To conclude this subsection, we present a numerical test for the validity of the CFT result for the logarithmic scaling of the R\'enyi entropy. 
Since the case of $A$ consisting of a single wire has been discussed and tested in Ref. \cite{cmv-12}, we focus here on a four-wire junction and the 
subsystem $A$ consisting of two wires. 
The $S$ matrix is chosen of the form
\begin{equation}
S=U\begin{pmatrix}
-\sqrt{1-s^2} & -s & 0 & 0\\
-s & \sqrt{1-s^2} & 0 & 0\\
0 & 0 & -1 & 0 \\
0 & 0 & 0 & 1 \\
\end{pmatrix}U^{-1} , \quad U=\begin{pmatrix}
1 & 0 & 0 & 0 \\
0 & -\cos \theta &  -\cos \theta \sin \theta & \sin^2 \theta \\
0 & \sin \theta & -\cos^2 \theta & \cos \theta \sin \theta\\
0 & 0 & \sin \theta & \cos \theta
\end{pmatrix}.
\end{equation}
The numerical results are reported in Fig. \ref{fig:n0} finding a perfect agreement with CFT.

\subsection{Entanglement negativity}

We now consider a tripartition $ A \cup B \cup C$, where $A$ ($B$) contains $M_{A}$ ($M_B$) wires, and we study the entanglement negativity between $A$ and $B$. This amounts to project the scattering-matrix $S$ over a subset of rows/columns belonging to $A \cup B$. In particular, we denote
\be
( C_{A\cup B})_{ij}(x,y) \equiv \la \Psi_j^\dagger(x)\Psi_i(y) \ra, \quad i,j =1,\dots, M_A+M_B,
\ee
as the correlation function of $A\cup B$, and
\be
\l S_{A\cup B} \r_{ij} \equiv S_{ij}, \quad i,j =1,\dots, M_A+M_B,
\ee
as the restriction of the scattering matrix;  $S_{A\cup B}$ is not unitary in general, and it satisfies the following relations
\be
(S_{A\cup B})^\dagger = (S_{A\cup B}), \quad 0\leq (S_{A\cup B})^2\leq 1.
\ee
Using the matrix representation of the correlation function $C$ in Eq. \eqref{eq_MatrixC} and restricting it to $A\cup B$, we obtain a $(2N(M_A+M_B),2N(M_A+M_B))$ 
matrix $C_{A\cup B}$.
The covariance matrix $\Gamma_{A\cup B}$ has the natural block form 
\be
\Gamma_{A\cup B}=  \begin{pmatrix} \Gamma_{AA} & \Gamma_{AB}  \\ \Gamma_{BA}  & \Gamma_{BB}\end{pmatrix},
\ee
from which we construct the matrix \cite{ssr-17,ez-15}
\be
\Gamma^\times_{A\cup B}\equiv \frac{2}{1+\Gamma^2_{A \cup B}} \begin{pmatrix} -\Gamma_{AA} & 0 \\ 0 & \Gamma_{BB}\end{pmatrix}.
\label{eq_GammaxAB}
\ee
The latter matrix $\Gamma^\times_{A\cup B}$ is the crucial object to write the R\'enyi negativities $\mathcal{E}_{n_e}$ which indeed are \cite{ssr-17}
\begin{multline}
\mathcal{E}_{n_e} \equiv \log \text{Tr}\l |\rho_{A\cup B}|^{n_e} \r =\\
 \text{Tr} \log\l \l\frac{1+\Gamma^\times_{A\cup B}}{2}\r^{n_e/2} + \l\frac{1-\Gamma^\times_{A\cup B}}{2}\r^{n_e/2} \r +\\ 
 \frac{n_e}{2}\text{Tr} \log\l \l\frac{1+\Gamma_{A\cup B}}{2}\r^{2} + \l\frac{1-\Gamma_{A\cup B}}{2}\r^{2} \r,
\label{eq_NegDefect}
\end{multline}
The above equation is valid for arbitrary real $n_e$ (i.e. also for a non-even integer)  
and so the negativity $\mathcal{E}$ is obtained just by taking  $n_e= 1$.

\begin{figure}[t]
\centering
	\includegraphics[width=0.5\linewidth]{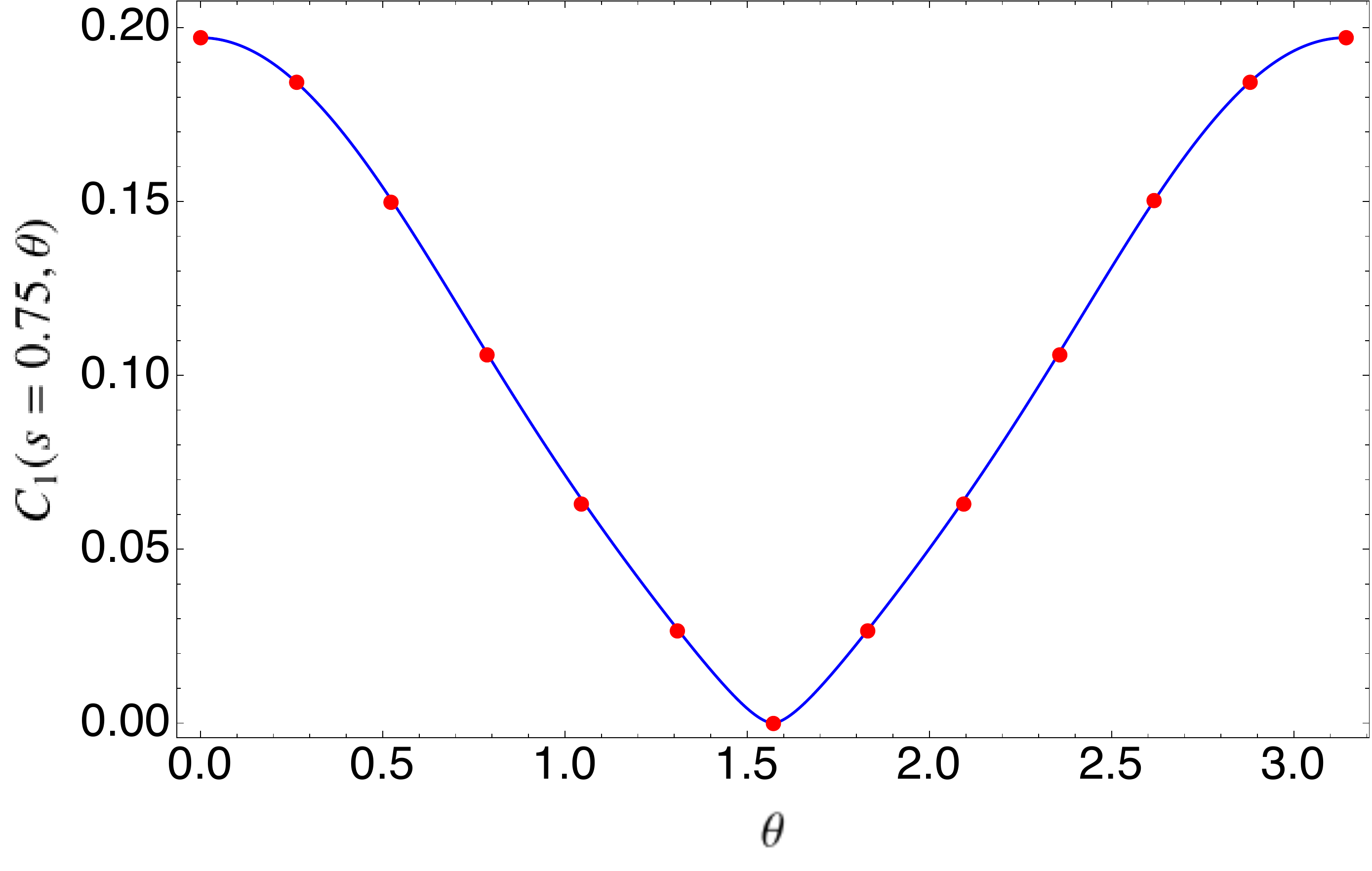}
    \caption{The coefficient of the logarithmic term of the negativity between two wires ($A$ and $B$) as a function of $\theta$, with fixed $s=0.75$. 
    The solid line corresponds to Eq. \eqref{eq:negativity} while the points have been obtained through a fit of the numerics with the form $a \log N+b_0+b_1 N^{-1}$. }
    \label{fig:n1}
\end{figure}

Eq. \eqref{eq_NegDefect} gives the R\'enyi negativities in terms of the correlation matrices that, once numerically evaluated, provides a test of the CFT 
results for the coefficient of the logarithm  obtained in Section \ref{sec:negativity}. 
For the numerical evaluation, we focus on a three-wire junction and on the two-parameter family of scattering matrices given by
\begin{equation}
S=U\begin{pmatrix}
-\sqrt{1-s^2} & -s & 0\\
-s & \sqrt{1-s^2} & 0 \\
0 & 0 & -1
\end{pmatrix}U^{-1} , \quad U=\begin{pmatrix}
1 & 0 & 0 \\
0 & -\cos \theta & \sin \theta \\
0 & \sin \theta & \cos \theta \\
\end{pmatrix}.
\end{equation}
We select as subsystems $A$ and $B$ the first two wires and compute numerically the R\'enyi negativity for several values of $s,\theta, $ and $N$.
In Fig. \ref{fig:n1} we reported the coefficient of the logarithm obtained as follows. 
We fixed $s=0.75$ and we selected some values of $\theta$;
for each value of $(s,\theta)$, we calculated numerically the negativity, for several values of $N$ up to 200. 
We fitted the obtained numerical results with $a \log N+b_0+b_1 N^{-1}$. 
Fig. \ref{fig:n1} finally reports the best fit of $a$ as a function of theta and compares it to the corresponding analytic result in Eq. \eqref{eq:negativity}, finding perfect agreement

\begin{figure}[t]
\centering
\subfigure
{\includegraphics[width=0.48\textwidth]{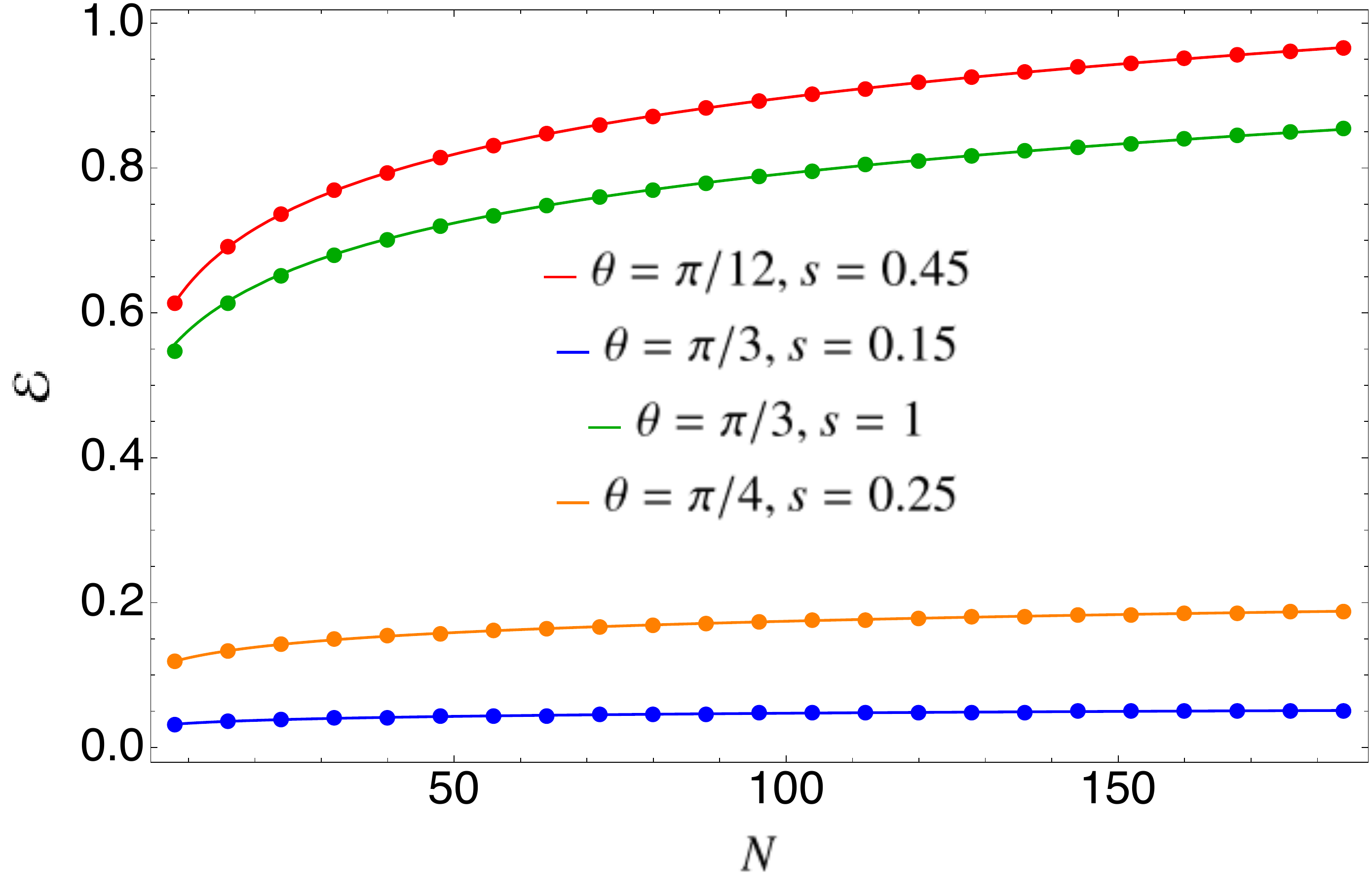}}
\subfigure
{\includegraphics[width=0.48\textwidth]{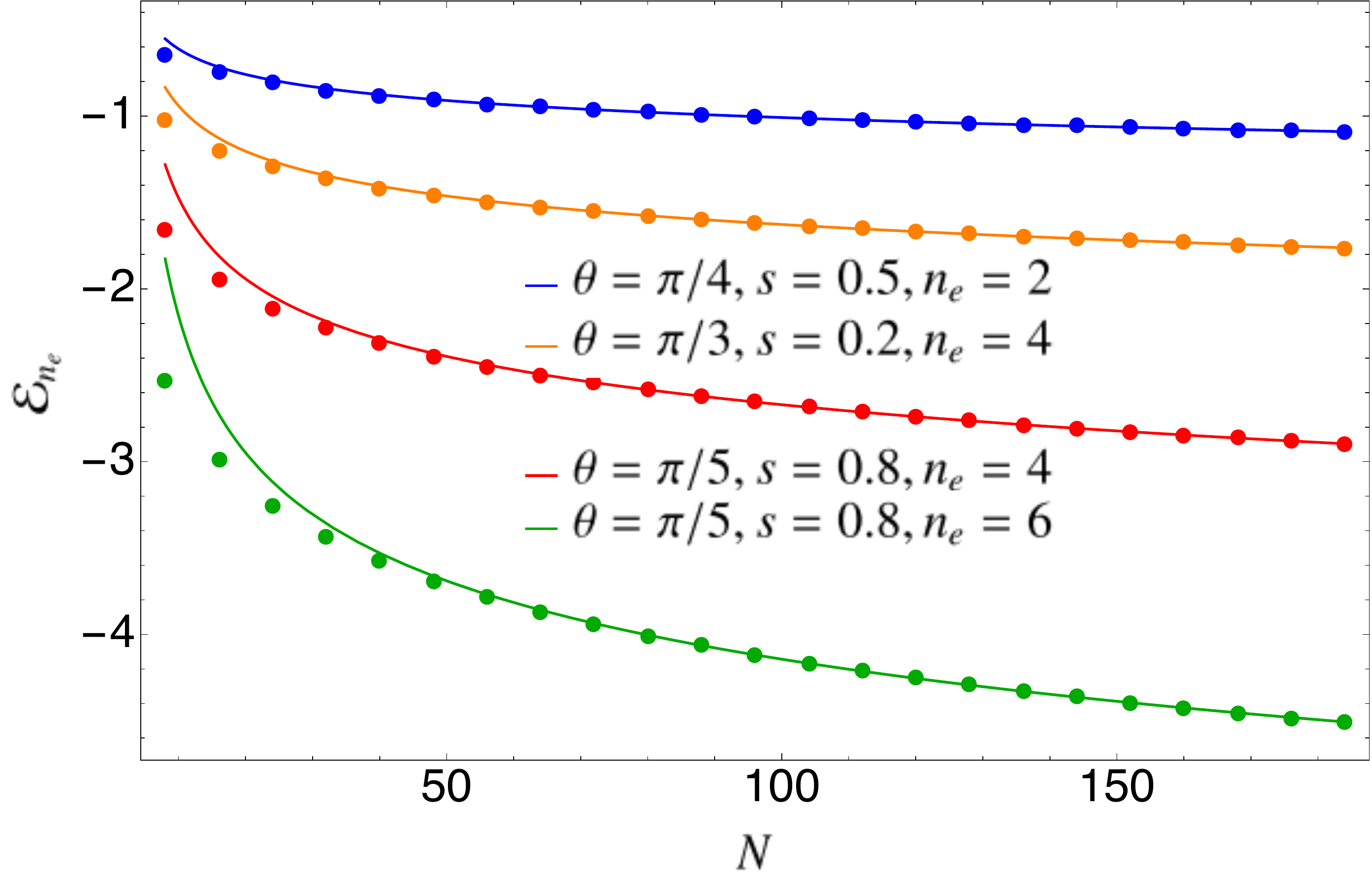}}
\caption{The R\'enyi negativities $\mathcal{E}_{n_e}$ in the three-wire junction for different values of $s,\theta$ and as a function of number of particles $N$. 
The lines show the curve $C_{n_e}(s,\theta)\log N+b_0+b_1N^{-1}$ where the coefficients $b_i$ are fitted using the data for $N \geq 80$. 
The prefactor $C_{n_e}(s,\theta)$ is given by Eq. \eqref{eq:negativity}.}\label{fig:n2}
\end{figure}

To conclude, in Fig. \ref{fig:n2} we report the $N$ dependence of the R\'enyi negativity and we benchmark the prefactor of the logarithmic term in 
Eq. \eqref{eq:negativity} for different pairs $(s,\theta)$ and different replica indices $n_e$.

\section{Conclusions}\label{sec:conlusion}
In this manuscript we investigated the entanglement entropy and negativity for the ground state of $M$ species of free massless Dirac fermions coupled at one boundary point 
via a conformal interface/junction. 
We consider a bipartion into two complementary sets with $M_A$ and $M_B$ wires each. 
We generalised the CFT approach (introduced in Ref. \cite{ss-08} for $M_A=M_B=1$) to evaluate the R\'enyi entanglement entropies for arbitrary $M_A$. 
In terms of the wire length $L$, the resulting R\'enyi entropy is 
\be
S_n(A)=C_n(S_{AA} )\log \frac{L}{\varepsilon}+O(1),
\ee
where $C_n(S_{AA})$ has been explicitly calculated and turns out to depend only on the scattering matrix reduced to the subsystem $A$, denoted by $S_{AA}$.
One interesting aspect of our result is that the $S_n(A)$ is  the sum of $M_A$ independent contributions which have the form of R\'enyi entropies of a single wire 
with a transmission probability given by the eigenvalues of $1-|S_{AA}|^2$ (see Eq. \eqref{eq:renyisum}). 
We obtained the same result also for a microscopical model of a free Fermi gas on a star junction. 
To do so, we generalised the approach of Ref. \cite{cmv-12} for $M_A=1$ to the case of a subsystem $A$ made of more wires.  
Such generalisation also provided numerical tests for the correctness of the CFT prediction.

We then moved to the study of the entanglement between two edges embedded in a multi-terminal junction, considering the fermionic negativity $\mathcal{E}$ of
Ref. \cite{ssr-17}. 
To this aim, we adapted the CFT formalism above to the computation of the R\'enyi negativities between two wires of a tripartite geometry.
As for the entanglement entropies, we found that the negativity grows logarithmically with the system size $L$, with a prefactor that 
depends on the details of the junction.
Also the negativity has been explicitly constructed and analysed for a free Fermi gas on a star junction, finding results fully compatible with the CFT.

We stress that our results apply to some different physical situations, too. 
For example, our predictions are expected to hold also for lattice models of free fermions, in particular for $M$ tight binding chains of length $L$ joined 
at a single common vertex (as, e.g., done in \cite{ep-12} for $M=2$). 
Furthermore, the logarithmic prefactors we obtained for entropy and negativity should appear also in the study of a star junction of $M$ infinite CFT, 
but when the subsystems consist of segments of length $\ell$ starting from the interface. 

We conclude the manuscript discussing few outlooks. 
The main focus of this work has been the free-fermion CFT, but our formalism can be easily adapted to free complex boson too, to study, e.g.,  
the entanglement entropy and negativity across junctions of harmonic chains.
The same formalism can be further applied to the study of some out-of-equilibrium protocols for free CFTs in the presence of defects (see, e.g., 
\cite{rfgc-22,coc-13,ep2-12,ge-20}).
Another interesting open problem is the generalisation to other CFTs such as the compact boson, WZW models  (see \cite{bbjs-16} for the interfaces of WZW),
or minimal models. 
In all those cases we do not expect any kind of replica diagonalisation, due to the lack of Gaussian measures, but still one could employ the replica construction
 to investigate the negativity, as well as other entanglement measures.
 
 \section*{Acknowledgements}
We are grateful to Mihail Mintchev for discussions. 
All authors acknowledge support from ERC under Consolidator grant number 771536 (NEMO).

\begin{appendix}

\section{Useful identities}\label{app1}
In this appendix  we report some technical details about the calculations appearing in Sections \ref{sec:CFTapproach} and \ref{sec:negativity}.

\subsection{Expectation value of Gaussian operators}

We want to prove the following identity for the expectation value of Gaussian operators
\be
\bra{0} \exp\l \mathcal{O}'_{jj'} \psi^{j}_{k} \bar{\psi}^{j'}_{k}\r\exp\l \mathcal{O}_{jj'} \bar{\psi}^j_{-k}\psi^{j'}_{-k}\r\ket{0} = \text{det}\l 1+\mathcal{O}'\mathcal{O} \r.
\label{eq_BoundOverlap}
\ee
Here $\psi^j_{\mp k}$ is the creation/annihilation operator of a Majorana fermion in the $k$-th left mode of the $j$-th species (among the $M$ ones), while $\bar{\psi}^j_{\mp k}$ is the corresponding right mover. $\mathcal{O}_{jj'}$ and $\mathcal{O}'_{jj'}$ are $M\times M$ matrices and the sum over $j,j'$ is implicit.

Before proving Eq. \eqref{eq_BoundOverlap} in the most general case, we  highlight simple cases in which it holds. Let us suppose there is just one species of fermions, so that we can suppress the indices $j,j'$ and the matrices $\mathcal{O}$ and $\mathcal{O}'$ become numbers. Using that the annihilation operators annihilate the vacuum, that the fermionic operators square to zero, and applying Wick theorem, one gets
\begin{multline}
\bra{0} \exp\l \mathcal{O}' \psi_{k} \bar{\psi}_{k}\r\exp\l \mathcal{O} \bar{\psi}_{-k}\psi_{-k}\r\ket{0} = \bra{0}\l 1+ \mathcal{O}' \psi_{k} \bar{\psi}_{k}\r \l 1+\mathcal{O} \bar{\psi}_{-k}\psi_{-k}\r\ket{0} = \\
\bra{0}1+\mathcal{O}'\mathcal{O} \psi_{k} \bar{\psi}_{k}\bar{\psi}_{-k}\psi_{-k}\ket{0} = 1+\mathcal{O}'\mathcal{O}.
\end{multline}
Similarly, when $\mathcal{O}$ and $\mathcal{O}'$ commute, one can diagonalise them simultaneously and apply the previous consideration to show Eq. \eqref{eq_BoundOverlap}.

We provide a general proof of Eq. \eqref{eq_BoundOverlap} using Gaussian integrals over Grassmann variables (whose basic properties  can be found on \cite{zj-89}). We start by representing the Gaussian operators as integrals over Grassmann variables. In particular, for each $j$-th Dirac fermionic field we associate a pair of Grassmann variables $\eta_j,\bar{\eta}_j$, and we express the Gaussian operator $\exp\l \mathcal{O}_{jj'} \bar{\psi}^j_{-k}\psi^{j'}_{-k}\r$ as follows
\be
\exp\l \mathcal{O}_{jj'} \bar{\psi}^j_{-k}\psi^{j'}_{-k}\r = \int d\eta d\bar{\eta}\exp\l -\eta_j\bar{\eta}_j +\eta_j \mathcal{O}_{jj'}\psi^{j'}_{-k} +\bar{\psi}^{j}_{-k}\bar{\eta}_j \r,
\label{eq:Gauss_rep1}
\ee
where the sums over $j$ and $j'$ are implicit. Similarly, we introduce a pair of Grassmann variables $\theta_j,\bar{\theta}_j$ to each species $j$ and we express  $\exp\l \mathcal{O}'_{jj'} \psi^{j}_{k} \bar{\psi}^{j'}_{k}\r $ as
\be
\exp\l \mathcal{O}'_{jj'} \psi^{j}_{k} \bar{\psi}^{j'}_{k}\r = \int d\theta d \bar{\theta}\exp\l -\bar{\theta}_j\theta_j + \psi^j_k \mathcal{O}'_{jj'}\theta_{j'}+\bar{\theta}_j\bar{\psi}^j_{k}\r.
\label{eq:Gauss_rep2}
\ee
Using the relations \eqref{eq:Gauss_rep1} and  \eqref{eq:Gauss_rep2} we write the product of the two Gaussian operators as follows
\be
\begin{split}
\bra{0} \exp\l \mathcal{O}'_{jj'} \psi^{j}_{k} \bar{\psi}^{j'}_{k}\r\exp\l \mathcal{O}_{jj'} \bar{\psi}^j_{-k}\psi^{j'}_{-k}\r\ket{0} = \int d\eta d\bar{\eta} d\theta d\bar{\theta} \exp\l
-\eta_j\bar{\eta}_j-\bar{\theta}_j\theta_j
\r\\
\bra{0}\exp\l \bar{\theta}_j\bar{\psi}^j_{k} \r \exp\l \bar{\psi}^{j}_{-k}\bar{\eta}_j\r\ket{0}\bra{0}\exp\l \psi^j_k \mathcal{O}'_{jj'}\theta_{j'}\r\exp\l \eta_j \mathcal{O}_{jj'}\psi^{j'}_{-k} \r\ket{0}=\\
\int d\eta d\bar{\eta} d\theta d\bar{\theta}\exp\l
-\eta_j\bar{\eta}_j-\bar{\theta}_j\theta_j+\bar{\theta}_j\bar{\eta}_j -\eta_j (\mathcal{O}\mathcal{O}')_{jj'}\theta_{j'} \r.
\end{split}
\label{eq:Grass_Integ}
\ee
The last step is the evaluation of the Gaussian integral over the $4M$ Grassmann variables $\eta_j,\bar{\eta}_j,\theta_j,\bar{\theta}_j$. We introduce a $4M$-dimensional vector $\Theta$ of Grassmann variables as follows
\be
\Theta = \begin{pmatrix}\eta \\ \bar{\theta} \\ \bar{\eta} \\ \theta \end{pmatrix},
\ee
and the Gaussian integral in \eqref{eq:Grass_Integ}
 as
 \be
\bra{0} \exp\l \mathcal{O}'_{jj'} \psi^{j}_{k} \bar{\psi}^{j'}_{k}\r\exp\l \mathcal{O}_{jj'} \bar{\psi}^j_{-k}\psi^{j'}_{-k}\r\ket{0} = \int d\Theta \exp\l -\frac{1}{2}\Theta^T \tilde{\mathcal{O}}\Theta\r,
\label{eq:Grass_int}
\ee
with $\tilde{\mathcal{O}}$ being the following $4M\times 4M$ matrix
\be
\tilde{\mathcal{O}} = \begin{pmatrix} 0 & 0 & 1 & \mathcal{O}\mathcal{O}' \\ 0 & 0 & -1 & 1 \\ -1 & 1 & 0 & 0\\ -(\mathcal{O}\mathcal{O}')^T & -1 & 0 & 0\end{pmatrix}.
\ee
The integral over $\Theta$ in Eq. \eqref{eq:Grass_int} gives the Pfaffian of the matrix $\tilde{\mathcal{O}}$ \cite{zj-89}, which we write as
\be
\text{Pf}\l \tilde{\mathcal{O}} \r = \text{det}^{1/2}\l \tilde{\mathcal{O}} \r = \text{det}\begin{pmatrix}  1 & \mathcal{O}\mathcal{O}' \\  -1 & 1 \end{pmatrix} = \text{det}(1+\mathcal{O}'\mathcal{O}).
\ee
This completes the proof of Eq. \eqref{eq_BoundOverlap}, which is the main result of this section.

\subsection{A useful determinant}

In this section we show that for a complex unitary $M\times M$ matrix  $S$ having the block structure \eqref{eq:S_block} the following relation holds
\begin{multline}
\text{det}\l \begin{pmatrix} 1 & 0 \\ 0 & 1\end{pmatrix}+ q^{2k}\begin{pmatrix} S^\dagger_{AA} & S^\dagger_{BA} \\ S^\dagger_{AB}  & S^\dagger_{BB}  \end{pmatrix} \begin{pmatrix} S_{AA} & e^{-i\alpha}S_{AB} \\ e^{i\alpha}S_{BA}  & S_{BB}  \end{pmatrix}\r =\\
 \text{det}\l 1 + 2(S_{AA}^\dagger S_{AA} + (1-S_{AA}^\dagger S_{AA})\cos \alpha )q^{2k} +q^{4k} \r  (1 + q^{2k})^{M - 2M_A}.
\label{eq:rem_det}
\end{multline}
Before proceeding with the proof, we notice that the $\alpha$ dependence in the rhs above is related only to the non-zero eigenvalues of $1-S^\dagger_{AA}S_{AA}$. Despite the explicit dependence of $1-S^\dagger_{AA}S_{AA}$ on $A$,
this fact does not lead to any asymmetry between $A$ and $B$. Indeed, the unitarity of $S$, $SS^\dagger = S^\dagger S = 1$, implies 
\be
S_{BB}S^\dagger_{BB} + S_{BA}S^\dagger_{BA} = 1, \quad S^\dagger_{AA}S_{AA}+ S^\dagger_{BA}S_{BA}=1.
\ee
Hence $1-S_{BB}S^\dagger_{BB} = S_{BA}S^\dagger_{BA}$ and $1-S^\dagger_{AA}S_{AA}= S^\dagger_{BA}S_{BA}$ have the same spectrum, up to zero eigenvalues, which means that \eqref{eq:rem_det} is symmetric by exchanging $A\leftrightarrow B$.

In order to prove Eq. \eqref{eq:rem_det}, we introduce a $M\times M$ matrix $\mathcal{O}$ as follows
\be
\begin{split}
\mathcal{O} = \begin{pmatrix} S^\dagger_{AA} & S^\dagger_{BA} \\ S^\dagger_{AB}  & S^\dagger_{BB}  \end{pmatrix} \begin{pmatrix} S_{AA} & e^{-i\alpha}S_{AB} \\ e^{i\alpha}S_{BA}  & S_{BB}  \end{pmatrix},
\end{split}
\ee
so that Eq. \eqref{eq:rem_det} requires the evaluation of  $\text{det}\l 1+q^{2k} \mathcal{O}\r$. 
Since, as a consequence of the unitarity of $S$,  $\mathcal{O}$ is unitary and it has the same spectrum of $\mathcal{O}^\dagger$, we can write
\be
\text{det}\l 1+ q^{2k}\mathcal{O} \r = \sqrt{\text{det}\l 1+ q^{2k}\mathcal{O} \r \text{det}\l 1+ q^{2k}\mathcal{O}^{\dagger} \r} = \sqrt{\text{det}\l 1+ q^{2k}(\mathcal{O}+\mathcal{O}^\dagger)+q^{4k} \r}.
\ee
Exploiting the unitarity of $S$, $\mathcal{O}+\mathcal{O}^\dagger$ has a block diagonal structure given by
\begin{equation}
    \mathcal{O}+\mathcal{O}^\dagger=\begin{pmatrix}
    2S^\dagger_{AA}S_{AA}+2(1-S^\dagger_{AA}S_{AA})\cos \alpha & 0 \\
    0 & 2S^\dagger_{BB}S_{BB}+2(1-S^\dagger_{BB}S_{BB})\cos \alpha
    \end{pmatrix}.
\end{equation}
Since we have already shown that $1-S^\dagger_{AA}S_{AA},1-S^\dagger_{BB}S_{BB}$ have the same non-zero spectrum, we get
\be
\text{det}\l 1+ q^{2k}\mathcal{O} \r \propto \text{det}\l 1+ q^{2k}(2S^\dagger_{AA}S_{AA}+2(1-S^\dagger_{AA}S_{AA})\cos \alpha)+q^{4k} \r,
\ee
The  $\alpha$-independent proportionality constant has to be a power of $(1+q^{2k})$, which comes form the possible presence of zero eigenvalues of 
$1-S^\dagger_{AA}S_{AA}$ (or $1-S^\dagger_{BB}S_{BB}$). 
We match this constant by power counting. More precisely, since $\text{det}\l 1+ q^{2k}\mathcal{O} \r$ is a polynomial in $q^{2k}$ of order $M$ and
\be
\text{det}\l 1+ q^{2k}(2S^\dagger_{AA}S_{AA}+2(1-S^\dagger_{AA}S_{AA})\cos \alpha)+q^{4k} \r
\ee
is a polynomial in $q^{2k}$ of order $2M_A$, the right power of $(1+q^{2k})$ which matches the proportionality constant has to be $(1+q^{2k})^{M-2M_A}$. This concludes the proof of Eq. \eqref{eq:rem_det}.

\subsection{Jacobi Theta functions and Dilogarithm }\label{sec_Jacobi}

Here we review some properties of the Jacobi Theta functions and the dilogarithm, useful for the evaluation of the partition functions. We consider the following infinite product representation for the Jacobi theta function $\theta_3(z,q)$ \cite{gr-94}
\be
\theta_3(z,q) = \prod_{m=1}^\infty 
\left( 1 - q^{2m}\right)
\left( 1 + 2 \cos(2 \pi z)q^{2m-1}+q^{4m-2}\right).
\ee
We want to evaluate $\theta_3(z,q)/\theta_3(0,q)$ in the limit $q\rightarrow 1^{-}$. To do so, we first take its logarithm, which turns the infinite product representation into a sum, i.e. 
\be
\log \frac{\theta_3(z,q)}{\theta_3(0,q)}  = \sum^{\infty}_{m=1}\left[\log(1+e^{i2\pi z} q^{2m-1}) + \log(1+e^{-i2\pi z} q^{2m-1}) -2\log(1+ q^{2m-1})\right].
\ee
For $q\simeq 1$, $q^{m}$ goes to zero slowly as $m$ grows and the sum can be approximated ($\simeq$) by an integral
\begin{multline}
\log \frac{\theta_3(z,q)}{\theta_3(0,q)} \simeq \int^{\infty}_{0}dx \ [\log(1+e^{i2\pi z}q^{1+2x})+\log(1+e^{-i2\pi z}q^{1+2x})-2\log(1+q^{1+2x})] =\\
-\frac{1}{2\log q}\int^{1}_{0}\frac{dt}{t}[\log(1+e^{i2\pi z}t)+\log(1+e^{-i2\pi z}t)-2\log(1+t)] =\\
\frac{1}{2\log q}\l \text{Li}_2(-e^{i2\pi z})+\text{Li}_2(-e^{-i2\pi z})-2\text{Li}_2(-1)\r = \frac{(2\pi z)^2}{4\log q}.
\end{multline}
In the previous computation, the integral representation of the dilogarithm function\cite{gr-94}
\be
\text{Li}_2(z) = -\int^{1}_0\frac{dt}{t}\log(1-zt)
\ee
has been employed, together with the property
\be
\text{Li}_2(-e^{i2\pi z})+\text{Li}_2(-e^{-i2\pi z}) = -\frac{\pi^2}{6}+\frac{(2\pi z)^2}{2},
\ee
valid for $z \in [-1/2,1/2]$.\\

\subsection{Trigonometric identities}\label{sec:trig}

We consider some useful algebraic identities which will be applied for the analytical continuation of the R\'enyi negativity. The first identity, which holds for any $n \in \mathbb{N}$ is
\be
\prod^{\frac{n-1}{2}}_{p=-\frac{n-1}{2}}(x+e^{i2\pi p/n}y) = x^n +y^n,
\ee
which can be proved through an explicit factorisation of the polynomial $x^n+y^n$ in the variable $x$. 
From the  previous identity one easily deduces (see also \cite{gr-94})
\be
\prod^{\frac{n-1}{2}}_{p=-\frac{n-1}{2}}(x^2+2\cos \frac{2\pi p}{n}xy + y^2) = \prod^{\frac{n-1}{2}}_{p=-\frac{n-1}{2}}(x+e^{i2\pi p/n}y)\prod^{\frac{n-1}{2}}_{p=-\frac{n-1}{2}}(x+e^{-i2\pi p/n}y) = (x^n+y^n)^2.
\label{eq:TrigId0}
\ee
The last non-trivial identity we need is
\be
\prod^{\frac{n_e-1}{2}}_{p=-\frac{n_e-1}{2}}\l x+ \cos \frac{2\pi p}{n_e} y\r = \l \l\frac{x+\sqrt{x^2-y^2}}{2}\r^{n_e/2} + \l\frac{x-\sqrt{x^2-y^2}}{2}\r^{n_e/2}\r^2,
\label{eq:TrigId}
\ee
which holds for even $n_e$. To prove Eq. \eqref{eq:TrigId}, we can write
\be
\prod^{\frac{n_e-1}{2}}_{p=-\frac{n_e-1}{2}}\l x+ \cos \frac{2\pi p}{n_e} y\r   =  \prod^{\frac{n_e-1}{2}}_{p=-\frac{n_e-1}{2}}\l x_1^2+2\cos \frac{2\pi p}{n_e}x_1y_1 + y_1^2\r ,
\ee
with $x_1$ and $y_1$ defined by
\be
\begin{cases} x_1^2+y_1^2 = x, \\ 2x_1y_1 = y\end{cases}.
\ee
The solution to the previous system is (up to $x_1\leftrightarrow y_1$)
\be
\begin{cases}x_1^2 = \frac{x+\sqrt{x^2-y^2}}{2},\\ y_1^2 = \frac{x-\sqrt{x^2-y^2}}{2}.  \end{cases}
\ee
If $n_e$ is even, we can write $x_1^{n_e} = \l x^2_1 \r^{n_e/2}$, replace $x_1$ as a function of $x,y$ (similarly for $y_1$), employ Eq. \eqref{eq:TrigId0}, obtaining \eqref{eq:TrigId}.\

We conclude this appendix with a straightforward consequence of \eqref{eq:TrigId}, which is the evaluation of the product
\be
\prod^{\frac{n_e-1}{2}}_{p=-\frac{n_e-1}{2}}\l x+ \cos \frac{2\pi p}{n_e} y + \cos^2 \frac{2\pi p}{n_e} z\r.
\ee
Since the term in the parenthesis is a second order polynomial in $\cos \frac{2\pi p}{n_e}$, it can be factorised in the two roots and 
Eq. \eqref{eq:TrigId} can be used for the computation of each of the two products in which it splits.

\end{appendix}

\end{document}